\documentclass[11pt,a4paper]{article}

\usepackage[truedimen,margin=34mm]{geometry} 

\usepackage{mathrsfs}
\usepackage{amssymb}
\usepackage{amsmath}
\usepackage{ascmac}
\usepackage{amsthm}
\usepackage[pdftex]{graphicx}
\usepackage[pdftex]{color}
\usepackage{booktabs}
\usepackage{setspace}
\usepackage[round]{natbib}
\usepackage{color}
\usepackage{url}
\usepackage{multirow}

\usepackage{titlesec}
\titleformat*{\section}{\large\bfseries}
\titleformat*{\subsection}{\it}

\newtheorem{thm}{Theorem}
\newtheorem{lem}{Lemma}
\newtheorem{cor}{Corollary}
\newtheorem{prp}{Proposition}

\def\al{{\alpha}}

\def\be{{\beta}}
\def\ga{{\gamma}}
\def\de{{\delta}}
\def\ep{{\varepsilon}}
\def\la{{\lambda}}
\def\si{{\sigma}}

\def\Ga{{\Gamma}}

\def\lah{{\widehat \la}}
\def\lat{{\widetilde \la}}

\def\E{{\rm E}}

\def\non{{\nonumber}}

\title{{\bf On Global-local Shrinkage Priors \\for Count Data}}

\date{}
\author{}

\begin{document}

\maketitle
\doublespacing

\vspace{-1.5cm}
\begin{center}
Yasuyuki Hamura$^1$, Kaoru Irie$^2$ and Shonosuke Sugasawa$^3$
\end{center}

\noindent
$^1$Graduate School of Economics, The University of Tokyo\\
$^2$Faculty of Economics, The University of Tokyo\\
$^3$Center for Spatial Information Science, The University of Tokyo

\vspace{5mm}
\begin{center}
{\bf \large Abstract}
\end{center}
Global-local shrinkage prior has been recognized as useful class of priors which can strongly shrink small signals towards prior means while keeping large signals unshrunk.
Although such priors have been extensively discussed under Gaussian responses, we intensively encounter count responses in practice in which the previous knowledge of global-local shrinkage priors cannot be directly imported.
In this paper, we discuss global-local shrinkage priors for analyzing sequence of counts.
We provide sufficient conditions under which the posterior mean keeps the observation as it is for very large signals, known as tail robustness property. 
Then, we propose tractable priors to meet the derived conditions approximately or exactly and develop an efficient posterior computation algorithm for Bayesian inference. 
The proposed methods are free from tuning parameters, that is, all the hyperparameters are automatically estimated based on the data. 
We demonstrate the proposed methods through simulation and an application to a real dataset.

\bigskip\noindent
{\bf Key words}: Heavy tailed distribution; Markov Chain Monte Carlo; Poisson distribution; Tail robustness

\newpage
%----------------------------------------------------------------------------%
%         Introduction 
%----------------------------------------------------------------------------%
\section{Introduction}

High-dimensional count data appears in a variety of scientific fields including genetics, epidemiology and social science. It is frequently observed in such data that many of those counts are very small and nearly zero except for some outliers. For example, in crime statistics where we divide the area of interest into small sub-regions, the number of occurrences of specific crime is likely to be small or zero in many sub-regions, while it is still important to detect ``hotspots'', i.e., the regions of the unexplained high crime rate. In this context, the Poisson-gamma model is obviously inappropriate, for the gamma prior shrinks all the observations uniformly, including the large signals that should be kept unshrunk, which might result in overlooking such meaningful regions. The desirable prior should account for both small and large signals and realize the flexible shrinkage effects on Poisson rates.

The prior of this type has been studied as global-local shrinkage prior for the Gaussian observations. The sparse signals of high-dimensional continuous observations are detected by the horseshoe prior, which exhibits the aforementioned property of shrinkage, being comparable to the variable selection \citep{carvalho2010horseshoe}. It is extended to the three-parameter beta distribution for more flexible modeling of sparseness \citep{armagan2011generalized}. 
In hierarchical models, such priors have been adopted for random effect distributions in small area estimation \citep{tang2018modeling} or default Bayesian analysis \citep{bhadra2016default}. 
For recent developments, see, for example, \cite{bhadra2019lasso} and the references therein.

While extensively studied for Gaussian data, the global-local shrinkage priors have not been fully developed for count data, although Poisson likelihood models with hierarchical structure are widely used in applications such as disease mapping (see, for example, \citealt{wakefield2006disease} and \citealt{lawson2013bayesian}). 
The theory related to the Poisson likelihoods has been well developed (e.g., \citealt{brown2013poisson} and \citealt{yano2019}), but not necessarily from the viewpoint of global-local shrinkage. 
The standard Bayesian models for count data is of Poisson-gamma type; the gamma prior for the Poisson rate shows the similarity to the global-local shrinkage prior if one assumes further hierarchical prior on the gamma scale parameters. 
In this context, the use of heavy-tailed hierarchical priors has already been practiced (e.g., \citealt{Zhu2019}), but the research on the general, statistical property of such priors has been limited. 
The theoretical properties of the Bayes estimators of those models have been investigated partially by \cite{datta2016bayesian} with the focus on (a generalized version of) the three-parameter beta prior for the analysis of zero-inflated count data. 
Our research is also concerned with the global-local shrinkage for count data, but especially from the rigorous viewpoint of heavy-tail property, which ensures the large signals are less or not at all affected by the shrinkage effect.

The objective of our research is to consider the effect of the hyperprior on the Bayes estimators (posterior means) of Poisson rates in terms of the robustness property. 
In doing so, we first define the concept of tail-robustness for the Bayes estimators mathematically. A robust Bayes estimator should keep large signals unshrunk, while retaining the strong shrink effect on small signals towards prior means, which is the {\it tail-robustness} we assess by our main theorem. In Section~\ref{sec:general}, Theorem~\ref{thm:asymptotic_bias} and Corollary~\ref{cor:ui} give sufficient conditions for the tail-robustness.

Requiring the tail-robustness for the Bayes estimators helps us restrict the class of priors we should use. 
The conditions in Theorem~\ref{thm:asymptotic_bias} reveal the importance of local shrinkage, or the individual scale parameter of gamma distribution customized for each Poisson rate, and support the use of two classes of hyperpriors proposed in Section~\ref{sec:proposed_prior}: the inverse-gamma prior and the newly-introduced extremely heavy-tailed prior. The inverse-gamma prior is a well-known distribution and easy to be integrated into the model. The asymptotic bias for large signals is shown to be negligible, hence the inverse-gamma prior is ``approximately'' tail-robust. The extremely heavily-tailed prior is a new class of probability distributions, whose density function is derived so as to satisfy the conditions for tail-robustness. In contrast to the inverse-gamma prior, this prior is exactly tail-robust. Both priors are conditionally conjugate for most of parameters in the model, which allows the fast and efficient posterior analysis by Gibbs sampler.

In the numerical study, we observe the properties of tail-robustness theoretically guaranteed for those priors, while the standard Poisson-gamma model suffers from the overly-shrunk Bayes estimators for outliers. The difference of two proposed priors are empirically confirmed in this numerical study; the inverse-gamma prior is better in the point estimations for small signals, having more shrinkage effect toward prior mean, while the extremely heavy-tailed prior is successful in quantifying the uncertainty for large counts, as shown in the coverage rates of posterior credible intervals. Despite this difference, both priors perform almost equally in the analysis of the actual crime data in Japan by detecting the hotspots of crimes that are overlooked in the Poisson-gamma models.

The rest of the paper is organized as follows. 
In Section~\ref{sec:general}, we consider theoretical argument regarding tail robustness and derive sufficient conditions for local priors to hold tail robustness.
In Section~\ref{sec:proposed_prior}, we propose two local priors and provide efficient posterior computation algorithms using Gibbs sampling. 
We also discuss some properties of the implied marginal priors and posteriors of Poisson rate. 
Section~\ref{sec:sim} is devoted to the numerical experiments for the extensive comparison of the proposed priors and other commonly-used priors/estimators under the various settings. The application to the real data of crimes in Tokyo metropolitan area, Japan, is discussed in Section \ref{sec:data}.
The step-by-step sampling algorithm is given in the Appendix, which would be useful for practitioners. 
Finally, R code implementing the proposed method is available at GitHub repository (\url{https://github.com/sshonosuke/GLSP-count}).

%----------------------------------------------------------------------------%
%         Global-local shrinkage prior
%----------------------------------------------------------------------------%
\section{Tail-robustness under count response}
\label{sec:general}

% backgrounds
\subsection{Hierarchical models for count data}

Our model has the following hierarchical representation; the $m$ observations $y_1,\dots , y_m$ are conditionally independent and modelled by, for $i=1,\ldots,m$,  
\begin{equation}\label{model}
y_i|\la_i\sim {\rm Po}(\eta_i\la_i), \ \ \ \la_i|u_i\sim {\rm Ga}(\alpha,\beta/u_i), \ \ \ u_i\sim \pi(u_i), 
\end{equation}
where $Po(\eta _i\la _i)$ is the Poisson distribution with rate $\eta _i\la _i$, and $Ga(\al , \be/u_i)$ the gamma distribution with shape $\al$ and rate $\be/u_i$, whose (conditional) mean is $\al / (\be / u_i)$. In addition, $\eta _i \in (0, \infty )$ is offset and known, $(\alpha ,\beta ) \in (0, \infty )^2$ are the hyperparameters, and $u_i \in (0, \infty )$ is a local scale parameter. 
The offset term, $\eta _i$, can be any known constant in general; in practice, it is flexibly modeled by regression with the log link function, as we examine in Section~\ref{sec:data}. In what follows, we assume $\eta _i=1$ for simplicity. The two rate parameters of the gamma prior, $\beta$ and $u_i^{-1}$, control the global and local shrinkage effects, respectively. Under this model, the Bayes estimator of Poisson rate $\la_i$ we consider is the posterior mean 
\begin{equation}
\begin{split}
\lat_i 
&= \E\Big[\frac{u_i}{\beta+u_i}(\alpha+y_i) \Big| y_i \Big] \text{,} \label{eq:postmean} \\
&= y_i-\E\Big[\frac{\be}{\be+u_i}\Big(y_i-\frac{\alpha u_i}{\be}\Big)\Big| y_i \Big] 
\end{split}
\end{equation}
where the expectation is taken with respect to the marginal posterior of $u_i$, so that the conditional posterior mean of $\la_i$ shrinks $y_i$ toward the prior mean $\al u_i / \be $.
Throughout the paper, we consider proper priors for $u_i$ only. 
The use of improper priors for $u_i$ results in the improper marginal of $\la_i$, and the posterior distribution of $\la_i$ would not successfully reflect the prior information, failing to shrink the Bayes estimator satisfactory.

\subsection{Tail-robustness of the posterior mean}
The appropriate choice of prior $\pi (u_i)$ is discussed in terms of the shrinkage effect realized in the Bayes estimator $\lat _i$. As stated in the introduction, the estimator should not be shrunk toward prior mean when the large signal is observed. This property is named as the tail-robustness \citep[e.g.][]{carvalho2010horseshoe}. 
The tail-robustness is mathematically defined as the property that 
    \begin{equation}
    \lim_{y_i \to \infty } | \lat _i - y_i |  = 0 \text{.} 
    \label{eq:tail_robustness}
    \end{equation}
This means that the (mean) absolute error loss tends to zero as $y_i \to \infty $. 
For fixed $u_i$, the Bayes estimator $( \al + y_i ) / (1 + \be / u_i )$ clearly loses the tail-robustness, which motivates the study of hierarchical prior for $u_i$. 
Throughout this paper, our primal interest is in this property defined in (\ref{eq:tail_robustness}), but we note that there have been other definitions of tail-robustness related to various loss functions. We discuss in details the difference of tail-robustness concepts in the Supplementary Material.

To consider the tail-robustness, the next theorem is useful in evaluating the asymptotic bias $\lim_{y_i \to \infty } ( \lat _i - y_i )$ for a variety of priors.

%  Theorem 
\begin{thm}
\label{thm:asymptotic_bias}
Assume that $\pi ( \cdot )$ is strictly positive and continuously differentiable. 
Suppose that $\pi ( \cdot )$ satisfies the following two conditions: 
\begin{align}
&\int_{0}^{1} |u \pi{}' (u)| du < \infty \text{,} \tag{A1} \label{A_1} \\
&\xi \equiv \lim_{u \to \infty } \frac{u {\pi }' (u) }{ \pi (u)} \quad \text{exists} \ \text{in} \  [- \infty , \infty ]  \text{.} \tag{A2} \label{A_2}
\end{align}
Then the asymptotic bias of $\lat _i$ is $1 + \xi $, that is, 
\begin{align}
\lim_{y_i \to \infty } ( \lat _i - y_i ) = 1 + \xi \text{.} \non
\end{align}
\end{thm}

The asymptotic bias of $\lat_i$ under $y_i\to\infty$ can be characterized by the tail behaviour of the mixing distribution $\pi(\cdot)$. This condition is similar to but significantly different from that of Gaussian response \citep[e.g.][]{tang2018modeling}. It is immediate from Theorem~\ref{thm:asymptotic_bias} that $\xi = -1$ is the sufficient condition for the estimator to be tail-robust, which is summarized in the following corollary.

%  Corollary 
\begin{cor}
\label{cor:ui}
Under the conditions (\ref{A_1}) and 
\begin{equation}
\lim_{u \to \infty } \frac{u {\pi }' (u) }{ \pi (u)} = -1\text{,} \tag{A3} \label{A_3}
\end{equation}
the Bayes estimator $\lat_i$ is tail-robust and satisfies $|\lat_i-y_i|\to 0$ as $y_i\to\infty$. 
\end{cor}

The crucial assumption in the above corollary is (\ref{A_3}), which describes the desirable tail behavior of the marginal prior distribution of $\la _i$. In fact, (\ref{A_3}) is sufficient for $\psi (u) = u \pi (u)$ to be slowly varying as $u \to \infty $, i.e., $\lim_{u \to \infty } \psi ( \kappa u ) / \psi (u) = 1$ for all $\kappa > 0$ (e.g., see \citealt[equation (1.11)]{senetaregularly}). It further implies that, for the marginal prior $p(\la _i) = \int _0^{\infty} Ga(\la _i|\al ,\be /u_i ) \pi (u_i) du_i$, we have $\la _i p(\la _i) \sim \la _i \pi (\la _i)$ 
as $\la _i \to \infty $ under the regularity condition that justifies the interchange of the limit and integral. In other words, under this assumption, the marginal densities of $\la _i$ and $u_i$ are asymptotically equivalent in the tail as density functions.

An example of priors that satisfies assumption (\ref{A_3}) is $\pi (u) \propto 1/u$. In many cases, (\ref{A_3}) requires priors to be of this from; see Section S4 of the Supplementary Material. 
However, this prior is improper. 
In other words, $\pi ( \cdot )$ have to be as heavy-tailed as improper priors for $\lat _i$ to be tail-robust. 
On the other hand, (\ref{A_1}) is merely a technical requirement for the proof.

One notable feature of Corollary \ref{cor:ui} is that the sufficient conditions for the tail-robustness, (\ref{A_1}) and (\ref{A_3}), are independent of the values of hyperparmeters $\al $ and $\be $. 
This setting about hyperparameters is a great contrary to other approaches, e.g., Proposition 1 of \cite{datta2016bayesian} where the tail-robustness is discussed for the limiting values of hyperparameters, i.e., $\be \to \infty $ or $\be \to 0$.

%----------------------------------------------------------------------------%
%        Global-local shrinkage priors
%----------------------------------------------------------------------------%
\section{Global-local shrinkage priors for count data}
\label{sec:proposed_prior}

%   Marginal prior distribution
\subsection{Proposed priors}
Under the hierarchical model (\ref{model}), we propose two families of priors for $u_i$. 
Each of them is indexed by a hyperparameter $\ga \in (0, \infty )$, which will be estimated in practice. 

The first prior is the inverse gamma (IG) prior given by 
\begin{align}
\pi_{\rm{IG}} ( u_i ; \ga ) = {\ga ^{\ga } \over \Ga ( \ga )} {1 \over {u_i}^{1 + \ga}} e^{- \ga / u_i} \text{,} \label{eq:prior-IG}
\end{align}
where $\ga > 0$. 
This is the density of the $\mathrm{IG}(\ga , \ga )$ distribution. 
It is clearly proper and conditionally conjugate, which simplifies the posterior computation by Markov chain Monte Carlo methods. 
From Theorem~\ref{thm:asymptotic_bias}, it holds that $\lim_{y_i \to \infty } ( \lat _i - y_i ) = - \ga$, indicating that the IG prior approximately satisfies the tail-robustness when $\gamma$ is small. 
The shape and rate parameters are identical in (\ref{eq:prior-IG}), so that we have $E[1/u_i]=1$, and the global parameter $\be $ can be interpreted as the marginal rate parameter of the gamma distribution for $\lambda_i$, i.e., the global shrinkage factor. 

Next, we newly introduce a conjugate prior. 
The extremely heavy-tailed (EH) prior is defined by the density 
\begin{align}
\pi _{\rm{EH}} ( u_i ; \ga ) =  {\gamma \over 1 + u_i} {1 \over \{ 1 + \log (1 + u_i ) \} ^{1 + \ga }} 
\label{eq:prior-log}
\end{align}
for $\ga > 0$. 
The EH prior can be seen as a modification of the scaled-beta prior; the details on the connection to the EH prior is discussed in the Supplementary Material. 
The additional logarithm function in (\ref{eq:prior-log}) contributes to the integrability of the density function. The use of log-term is often seen in the literature of decision-theoretic statistical theory (for example, see \citealt[Remark 4.1]{maruyama2019admissible}). This prior is proper because 
\begin{align}
\int_{0}^{\infty } \pi _{\rm{EH}} (u; \ga ) du &= \Big[ - \{ 1 + \log (1 + u) \} ^{- \ga } \Big] _{0}^{\infty } = 1 \text{.} \non
\end{align}
The notable property of the EH prior is that it satisfies the condition of Corollary~\ref{cor:ui}; 
\begin{align}
{u {\pi_{\rm{EH}}}' (u; \ga ) \over \pi _{\rm{EH}} (u; \ga )} = u \Big\{ - {1 \over 1 + u} - {1 + \ga \over 1 + \log (1 + u)} {1 \over 1 + u} \Big\} \to - 1 \non
\end{align}
as $u \to \infty $. Hence, the EH prior is exactly tail-robust.

The densities and tail-behaviors of the proposed priors are summarized in Table~\ref{tab:prior} together with those of the Gauss hypergeometric (GH) prior considered in \citet{datta2016bayesian}. 
The GH prior is dependent on the global rate parameter $\beta$, but its density tail (the asymptotic functional form of density as $u_i\to \infty$) is independent of $\beta$ and identical to that of the half-Cauchy prior \citep{carvalho2010horseshoe}. 
The density tail of the EH prior is heavier than those of the GH and IG priors regardless of $\gamma$. This difference originates from the log-term of the EH density and contributes to the exact tail-robustness of the EH prior. 

\begin{table}[htp!]
	\begin{center}
		\begin{tabular}{lcc} \toprule 
			& Density kernel of $u_i$ &  Density tail as $u_i\to \infty$ \\ \midrule  
			$GH(1/2,1/2,\gamma, 1/\beta)$ & $u_i^{-1/2} (1+u_i)^{-\gamma} (\beta + u_i)^{\gamma -1}$  & $u_i^{-3/2}$ \\
			$IG(\gamma ,\gamma )$ & $u_i^{-(\gamma +1)} e^{-\gamma /u_i}$ & $u_i^{-(\gamma +1)}$ \\ 
			$EH(\gamma )$ & $(1+u_i)^{-1} \{ 1+\log(1+u_i) \}^{-(1+\gamma)}$ & $u_i^{-1} (\log u_i)^{-(1+\gamma )} $ \\ \bottomrule
		\end{tabular}
	\end{center}
	\caption{Densities of GH, IG and EH priors} \label{tab:prior}
\end{table}

Finally, we note the parametrization by $\kappa = 1 / (1 + u) \in (0, 1)$, which also clarifies the difference of the proposed priors from others. The implied density of the EH prior in the scale of $\kappa$ is $\pi _{EH} (\kappa ) = \ga \kappa ^{- 1} / \{ 1 + \log (1 / \kappa ) \} ^{1 + \ga }$. 
This expression shows that the EH prior can be viewed as an extension of the improper beta prior, $Be(0,1)$. The resulting EH prior is proper; the additional log-term in the density of the EH prior ensures the propriety. 
Other class of priors, including the half-Cauchy prior, remain in the class of beta distributions in $\kappa$-scale and do not involve the log-term in their densities.

%   Posterior computation
\subsection{Posterior computation}
\label{subsec:posterior_computation}

The computation of the Bayes estimator is based on the Markov chain Monte Carlo method. 
Because the proposed priors are mostly conditionally conjugate, sampling from most of the conditional posterior distributions is straightforward. 
In this subsection, we mention the essential strategies of the sampling methods.  We provide the detailed step-by-step Gibbs sampling in the Appendix.

We first discuss the parameters $(\la _{1:m}, \alpha ,\beta )$, which are common to and always included in all the models regardless of the choice of prior for $u_i$. Note that we assign prior distributions for $\al $ and $\be $ in practice. 
In this research, we consider the gamma priors; $ \alpha \sim {\rm{Ga}} ( a_{\al } , b_{\al } )$ and $\be \sim {\rm{Ga}} ( a_{\be } , b_{\be } )$. 
We adopt $a_{\al }=b_{\al }=a_{\be }=b_{\be }=1$ as default choices, which will be used in the numerical studies in Sections \ref{sec:sim} and \ref{sec:data}.
When the model is of Poisson-gamma type and the local parameters $u_i$ are fixed, the posterior analysis can be done by sampling the above parameters. 
It is noted that the gamma prior for $\beta$ is conditionally conjugate whereas the gamma prior for $\alpha$ is not.
However, using the augmentation technique by \cite{zhou2013negative}, we can derive an efficient Gibbs sampling method as provided in the Appendix.

For the model with the IG prior, the scale parameter $u_i$ has a known conditional posterior, while the conditional posterior of the hyperparameter $\ga$ is difficult to directly sample from. 
Although several computationally-sophisticated options are available for the sampling of $\ga$, we here simply use the random-walk Metropolis-Hastings method with uniform prior $\gamma\sim U(\ep_1,\ep_2)$ for fixed small $\ep_1>0$ and large $\ep_2>0$.
We set $\ep_1=0.001$ and $\ep_2=150$ as a default choice.

The new EH prior is not conditionally conjugate for $u_i$, despite its simple closed-form of the density function in (\ref{eq:prior-log}). 
To develop an efficient sampling algorithm, we introduce a novel augmentation approach using two positive valued latent variables $v_i$ and $w_i$, given by  
\begin{align}
\pi _{\rm{EH}} ( u_i ; \ga ) = \iint_{(0, \infty )^2} \pi _{\rm{EH}} ( u_i , v_i , w_i; \ga ) d{v_i} d{w_i} \text{,} \non 
\end{align}
where 
\begin{align*}
\pi _{\rm{EH}} ( u_i , v_i , w_i; \ga ) &= Ga(u_i | 1,v_i) Ga(v_i|w_i,1) Ga(w_i|\ga ,1) \\ 
&= \frac{w_i^{\gamma-1}v_i^{w_i}}{\Gamma(\gamma)\Gamma(w_i)} \exp\left\{- w_i - v_i(1 + u_i )\right\} \text{.} \non 
\end{align*}
Using the above expression, it is observed that the full conditional distribution of $u_i$ is the generalized inverse Gaussian (GIG) distribution. 
We can also obtain familiar forms of the conditional posterior distributions of the other parameters, $(v_i,w_i)$, where the details are given in the Appendix. 
For the shape parameter $\gamma$ in the EH prior, we assign gamma prior $\gamma\sim {\rm Ga}(a_{\ga},b_{\ga})$ which is conditionally conjugate. 
We use $a_{\ga}=b_{\ga}=1$ for simplicity as a default choice.

%   Marginal prior distribution
\subsection{Marginal prior distributions for $\la_i$}
\label{subsec:marginal}
In this section, the marginal density of $\la_i$ is computed to consider its behavior in the limit of $\la _i \to \infty$ and $\la _i\to 0$. 
The tail property of the marginal density is the same as that of the prior of $u_i$. 
Information on the behavior of the marginal density of $\la _i$ around zero is also important to understand the amount of shrinkage effect toward zero, which has not been discussed up to this point in this paper. 
In general, the marginal prior distribution for $\la _i$ is given by 
\begin{align}
p( \la _i ; \al , \be , \ga ) 
&= \int_{0}^{\infty } {\be ^{\al } / {u_i}^{\al } \over \Ga ( \al )} {\la _i}^{\al - 1} e^{- ( \be / u_i ) \la _i} \pi ( u_i; \ga ) d u_i \non \\
&= {\be ^{\al } \over \Ga ( \al )} \int_{0}^{\infty } {1 \over x^{\al }} e^{- \be / x} \pi ( \la _i x; \ga ) dx \text{.} \non 
\end{align}
We continue the computation of this density for the two classes of priors: $\pi _{IG}$ and $\pi _{EH}$. 

For the IG prior $\pi ( u_i ; \ga ) = \pi_{\rm{IG}} ( u_i ; \ga )$, we have 
\begin{align}
p( \la _i ; \al , \be , \ga ) = {( \be / \ga )^{\al } \over B( \al , \ga )} {{\la _i}^{\al - 1} \over \{ 1 + ( \be / \ga ) \la _i \} ^{\al + \ga }} \text{,} \label{eq:marginal_IG}
\end{align}
which implies the beta distribution, i.e., 
\begin{align}
{( \be / \ga ) \la _i \over 1 + ( \be / \ga ) \la _i} \sim {\rm{Beta}} ( \al , \ga ) \text{.} \non
\end{align}
From (\ref{eq:marginal_IG}), we have $p( \la _i ; \al , \be , \ga ) = O( {\la _i}^{- 1 - \ga } )$ as $\la _i \to \infty $. For sufficiently small $\ga$, the marginal prior of $\la_i$ can be heavily-tailed, being almost equivalent to $\la _i^{-1}$ in the tail. This observation is coherent with the $\ga$-dependent asymptotic bias of the Bayes estimator, $\lim_{y_i \to \infty } ( \lat _i - y_i ) = - \ga $.  
It should also be noted here that, due to the heavy tail of this density, the prior mean does not exist if $\ga \le 1$. This can be confirmed by the direct computation or the fact that the prior mean of $u_i$ is not finite under this condition. In this situation, it is difficult to interpret the prior from the viewpoint that the estimator is shrunk toward the prior mean. For those who prefer the prior with finite mean, we recommend the modification of the IG prior to ${\rm{IG}} ( \ga + 1, \ga )$, $\ga > 0$, which instead increases the asymptotic bias slightly to $- \ga - 1$.

In contrast, the density at the origin depends on the value of $\al$. In particular, $\lim_{\la _i \to 0} p( \la _i ; \al , \be , \ga ) = \infty$ for $\al < 1$, while the limit becomes a positive constant for $\al = 1$ and zero for $\al > 1$. This fact gives a clue to the interpretation of the choice of, or the posterior inference for, hyperparameter $\al$.

For the EH prior, the marginal density is evaluated around zero as follows: 
For $\pi ( u_i ; \ga ) = \pi_{\rm{EH}} ( u_i ; \ga )$,  
\begin{align}
p( \la _i ; \al , \be , \ga ) &= {\be ^{\al } \ga \over \Ga ( \al )} \int_{0}^{\infty } {e^{- \be / x} \over x^{\al }} {1 \over 1 + \la _i x} {1 \over \{ 1 + \log (1 + \la _i x) \} ^{1 + \ga }} dx \non \\
&\to {\be ^{\al } \ga \over \Ga ( \al )} \int_{0}^{\infty } {e^{- \be / x} \over x^{\al }} dx \non \\
&= \begin{cases} (\al-1)^{-1}\be \ga & \text{if $\al > 1$} \\ \infty & \text{if $\al \le 1$} \end{cases} \non
\end{align}
as $\la _i \to 0$ by the monotone convergence theorem. 
Thus, $\lim_{\la _i \to 0} p( \la _i ; \al , \be , \ga ) > 0$ and increasing as $\alpha \to 0$, implying the shrinkage of small signals toward the global prior mean. 
For the tail property, we have
\begin{align}
\lim_{\la _i \to \infty } {p( \la _i ; \al , \be , \ga ) \over \pi_{\rm{EH}} ( \la _i ; \ga )} &= {\be ^{\al } \over \Ga ( \al )} \int_{0}^{\infty } {e^{- \be / x} \over x^{\al }} \Big[ \lim_{\la _i \to \infty } {1 + \la _i \over 1 + \la _i x} \Big\{ {1 + \log (1 + \la _i ) \over 1 + \log (1 + \la _i x)} \Big\} ^{1 + \ga } \Big] dx \non \\
&= {\be ^{\al } \over \Ga ( \al )} \int_{0}^{\infty } {e^{- \be / x} \over x^{\al + 1}} dx = 1 \text{.} \non 
\end{align}
Therefore, $p( \la _i ; \al , \be , \ga ) \sim \pi_{\rm{EH}} ( \la _i ; \ga ) \sim \ga {\la _i}^{- 1} ( \log \la _i )^{- 1 - \ga }$ as $\la _i \to \infty $, which means that the marginal prior $p( \la _i ; \al , \be , \ga )$ is proper but has a sufficiently heavy tail so that the model can accommodate large signals. For the computation to verify the result above, see the Supplementary Material.

\begin{figure}[!htbp]
	\centering
	\includegraphics[width=13cm]{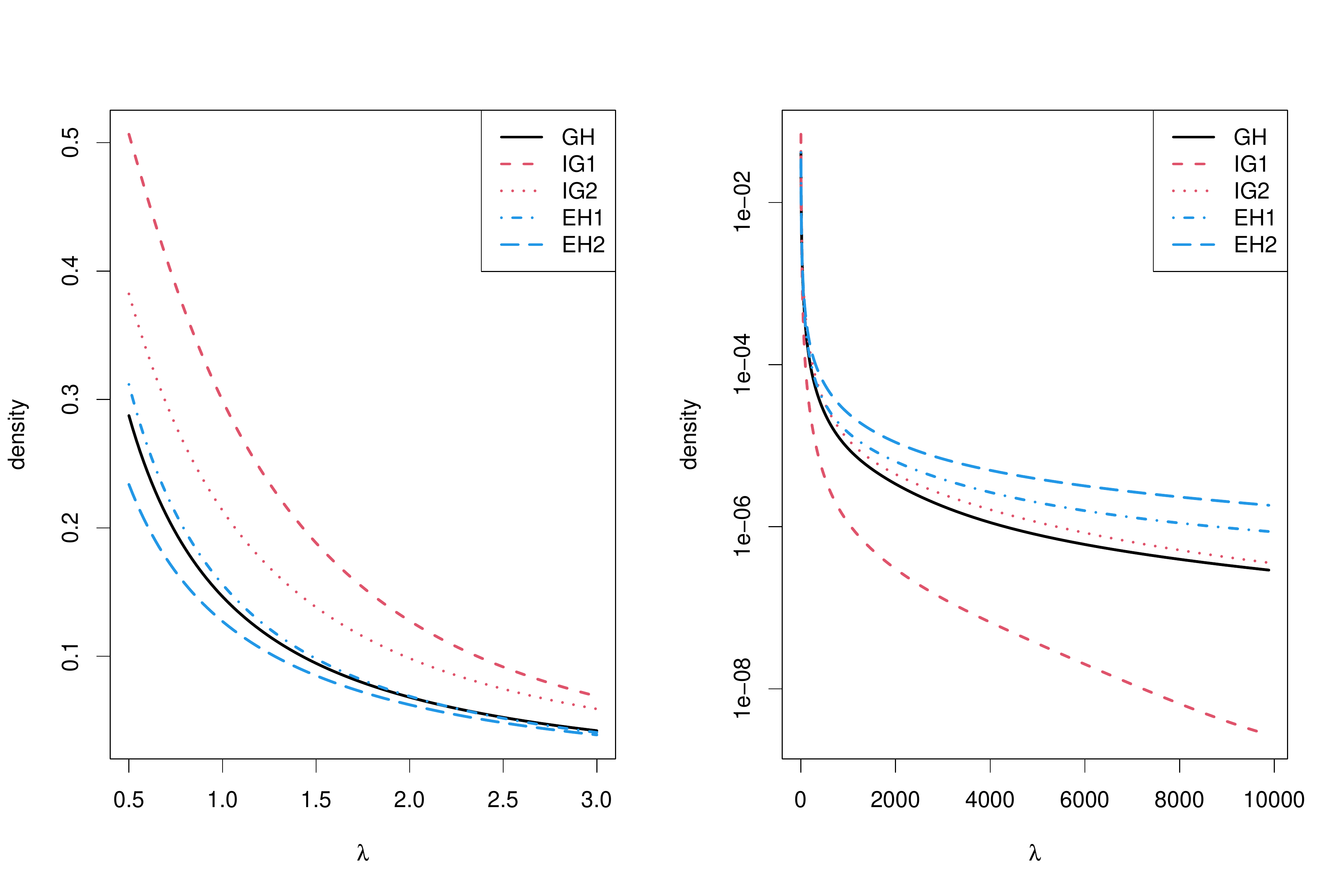}
	\caption{\small Left: Marginal densities of $\la_i$ with $\alpha=\beta=2$ under the Gauss hypergeometric prior (GH) with $\gamma=1$, inverse-gamma priors with $\gamma =1$ (IG1) and $\gamma = 0.5$ (IG2), and extremely heavily-tailed priors with $\gamma=1$ (EH1) and $\gamma = 0.5$ (EH2). The GH and EH densities are evaluated by the Monte Carlo integration. Right: The marginal densities of the five prior distributions in the tail. The vertical axis is logarithmic.
	} 
	\label{fig:prior}
\end{figure}

The marginal distributions of $\la_i$ with $\alpha=\beta=2$ under the proposed IG and EH priors with $\gamma=1$ and $\gamma=0.5$ as well as the GH prior with $\ga = 1$ are visually illustrated in Figure~\ref{fig:prior}.
It shows that the IG prior with $\gamma=0.5$ has almost the same tail-behavior as the GH prior since the tail-behavior of the density of $u_i$ under the IG prior with $\gamma=1$ is equivalent that of GH as confirmed in Table \ref{tab:prior}.
Moreover, the figure reveals that the density tail under the EH prior is heavier than those under the IG and GH priors, which is consistent with Table \ref{tab:prior}.

%  Figure

%  Posterior distribution 
\subsection{Marginal posterior distributions for $\la_i$}
\begin{figure}[!htbp]
	\centering
	\includegraphics[width=13cm]{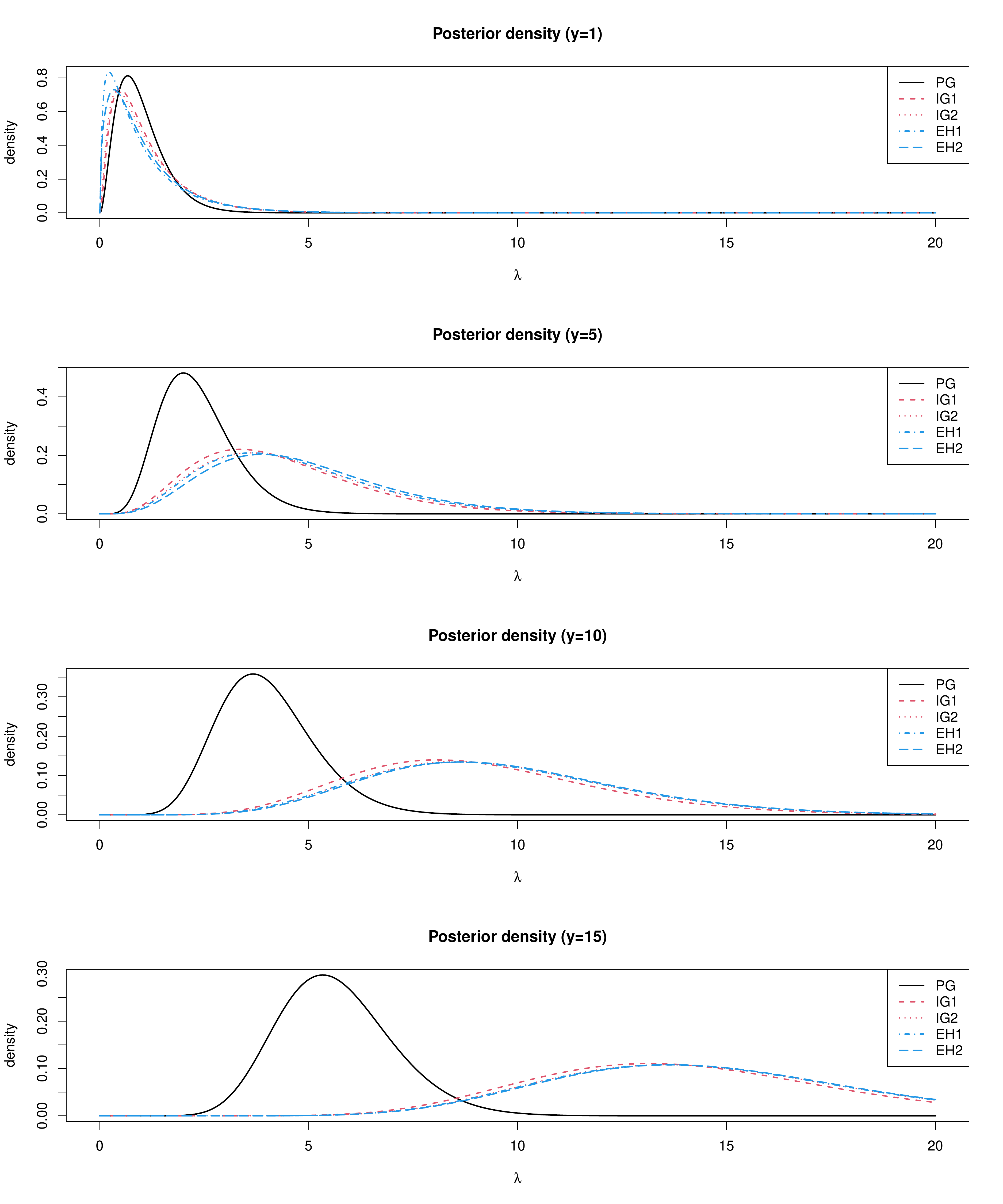}
	\caption{
	Marginal posterior distributions for $\la_i$ with $\alpha=\beta=2$ based on the conventional gamma prior (PG), the proposed inverse gamma prior with $\gamma=1$ (IG1) and $\gamma=0.5$ (IG2), and the proposed extremely heavy-tailed prior with $\gamma=1$ (EH1) and $\gamma=0.5$ (EH2).
Each row corresponds to a difference value of $y_i\in \{1,5,10,15\}$. 
	 \label{fig:post}
	 }
\end{figure}

We briefly describe the flexibility of the proposed prior distributions compared with the common gamma prior for $\la_i$.
Since the conditional posterior distribution of $\la_i$ given $u_i$ is ${\rm Ga}(y_i+\alpha,1+\beta/u_i)$ under the model (\ref{model}), the marginal posterior distribution of $\la_i$ is obtained as the mixture of the gamma distribution with respect to the marginal posterior distribution of $u_i$.
Note that the use of the gamma prior distribution for $\la_i$ leads to the posterior distribution ${\rm Ga}(y_i+\alpha,1+\beta)$.
We set $\alpha=\beta=2$ and show the marginal posterior density of $\la_i$ with several values of $y_i$ in Figure \ref{fig:post}.
It is observed that under the moderate signal such as $y_i=1$, the posterior distributions of $\la_i$ are almost the same among the conventional gamma prior and the proposed global-local shrinkage priors.
On the other hand, under large values of $y_i$, the posterior densities of the proposed methods are significantly different from one based on the gamma prior, which shows the flexibility of the proposed priors against large signals and is consistent with tail-robustness property given in Theorem \ref{thm:asymptotic_bias}.
However, the posterior density with the conventional gamma prior is not sensitive to large signals, which leads to the over-shrinkage of estimators.
As noted in the previous section, the hyperparameter $\gamma$ in the inverse gamma (IG) distribution is directly related to the asymptotic bias, and Figure \ref{fig:post} shows that the IG prior with the smaller $\gamma$ produces heavier-tailed posterior density functions than that with the larger $\gamma$.

% figure

%----------------------------------------------------------------------------%
%         Simulation 
%----------------------------------------------------------------------------%
\section{Simulation study}
\label{sec:sim}
We here investigate the finite sample performance of the proposed method together with some existing methods.
We generated the independent sequence of counts from $y_i\sim {\rm Po}(\la_i\eta_i)$ for $i=1,\ldots,m$ with $m=200$.
The adjustment term $\eta_i$ was generated from $U(1, 5)$, and assumed to be known. 
For the generating process for $\la_i$, we considered the mixture: $\la_i\sim (1-\omega) f_0+\omega f_1$, where $f_0$ and $f_1$ denote distributions of moderate and large signals, respectively.
Note that $\omega$ denotes the proportion of large signals (outliers).
For the settings of $f_0$ and $f_1$, we adopted the following four scenarios:
\begin{align*}
{\rm (I)}& \ \ \  f_0={\rm Ga}(2,2), \ \ \ f_1={\rm Ga}(10,2)\\
{\rm (II)}& \ \ \  f_0=0.75{\rm Ga}(2,2)+0.25\delta(1), \ \ \ f_1={\rm Ga}(10,2)\\
{\rm (III)}& \ \ \  f_0=0.5{\rm Ga}(2,2)+0.5\delta(1), \ \ \ f_1={\rm Ga}(10,2)\\
{\rm (IV)}& \ \ \  f_0=U(0,2), \ \ \ f_1=4+|t_3|,
\end{align*}
where $U(0,2)$ is the uniform distribution on $[0,2]$ and $t_3$ is the $t$-distribution with $3$ degrees of freedom. 
In scenarios (II) and (III), the moderate signals are more concentrated around 1 and have less variation, in comparison to the continuous prior ${\rm Ga}(2,2)$ in scenario (I). 
We define the outlying and non-outlying values of $\lambda _i$'s as those generated from $f_1$ and $f_0$, respectively. 
In each scenario, we considered two scenarios of $\omega$, namely, $\omega=0.05$ and $0.1$.

We considered the estimation of $\la_i$ using the following six priors/methods: 
\begin{itemize}
\item 
IG: The proposed method with inverse gamma prior for $u_i$.

\item
EH: The proposed method with extremely heavy tailed prior for $u_i$.

\item
GH: Gauss hyper-geometric prior proposed by \cite{datta2016bayesian}.

\item
PG: Using gamma distribution for $\la_i$, known as Poisson-gamma model.

\item
KW: Nonparametric empirical Bayes method 
(\citealt{KW1956}; \\ \citealt{koenker2014convex}).

\item
ML: Maximum likelihood (non-shrinkage) estimator, i.e., $y_i$.
\end{itemize}

We assigned prior distributions for the hyperparameters in the two proposed methods, as illustrated in Section \ref{subsec:posterior_computation}.
In the GH method, the hyperparameters were estimated by the empirical Bayes method recommended in \cite{datta2016bayesian}, and then 3,000 posterior samples were generated directly from the posterior distribution of $\lambda_i$ with the estimated hyperparameters. 
We assigned gamma priors for the hyperparameters in the PG method, and used the prior distributions given in Section \ref{subsec:posterior_computation} for the hyperparameter in the IG and EH methods.
The three methods require the computation by Markov chain Monte Carlo method; for each dataset, we generated 3,000 posterior samples after discarding 500 samples as a burn-in period.  
We computed point estimates of $\lambda_i$, where we used the posterior mean as point estimation in the first four methods. 
The performance of these point estimators are evaluated by the mean squared errors (MSE) and mean absolute percentage error (MAPE) defined as the averaged values of $(\lah_i-\la_i)^2$ and $|\lah_i-\la_i|/\la_i$, respectively. 
These measures were calculated separately for outlying and non-outlying values of the true $\lambda_i$'s. 
We also computed $95\%$ credible intervals of $\lambda_i$ based on the first four Bayesian methods, and evaluate the performance using the coverage probability (CP) and average length (AL). 
We repeated the process for 1,000 times to report the averages of MSE, MAPE, CP and AL below.

In Table \ref{tab:sim}, we presented the averaged values of the MSEs and MAPEs in all the scenarios. 
For non-outlying values, we can see that the PG and KW methods perform quite well while the proposed IG method is quite comparable. 
For non-outlying values, the performance of the three methods, IG, KW and ML are quite comparable and better than the other methods in MSE.
However, it should be noted that the EH performs best in MAPE.
These results would show that the shrinkage effects of the proposed methods successfully realized for small signals. 
On the other hand, for outlying values, the point estimates of both PG and KW methods tend to be worse than ML as predicted theoretically; the PG and KW methods are not tail-robust in general and are expected to produce over-shrunk estimates. 
In contrast, the proposed IG and EH methods as well as the GH method provides better performance than the PG and KW methods for outliers, as designed. Among the three methods, the EH method provides the best performance in all experiments, which is consistent with the tail-robustness property of the EH method, as discussed in Section~\ref{sec:proposed_prior}, noting that the IG and GH methods does not necessarily hold the property.

In Table \ref{tab:sim-CI}, we reported averaged values of the CPs and ALs of $95\%$ credible intervals of the four Bayesian methods. 
It is observed that all the method provides reasonable CP for non-outlying values whereas the CP of the PG method is seriously smaller than the nominal level for outlying values, which also shows the serious over-shrinkage property of the PG method. 
On the other hand, the proposed methods and the GH method show much higher CPs, while the CP of the EH method is much closer to the nominal level than that of the IG method.
It is also observed that the performance of the EH and GH methods are quite comparable in both CP and AL.

We checked the performance of the Markov chain Monte Carlo sampling algorithm for the IG, EH and IG methods under scenario (I) with $\omega=0.1$.
The averaged values of the inefficiency factors of $\lambda_1,\ldots,\lambda_m$ under the IG, EH and PG methods were $1.17$, $4.39$ and $1.01$, respectively.  
It shows that the resulting inefficiency factors seems acceptable, but that of the EH method is slightly higher than those of the other methods possibly because the number of latent parameters used in the Gibbs sampling of the EH method is large compared with the other methods.
In the Supplementary Material, we report the additional simulation studies with large sample size, namely, $m=400$, and computation time of the four Bayesian methods.

%  Table
\begin{table}[htbp!]
\caption{Averaged values of mean squared errors (MSE) and mean absolute percentage error (MAPE) in non-outlying (-n) and outlying (-o) areas under four scenarios with $m=200$ and $\omega\in \{0.05, 0.1\}$.  
\label{tab:sim}
}
\begin{center}
\begin{tabular}{ccccccccccccccc}
\hline
Scenario  & $\omega$ &  &  & IG & EH & GH & PG & KW & ML \\
\hline
\multirow{4}{*}{(I)}  &  \multirow{4}{*}{0.05} & MSE-n &  & 0.24 & 0.28 & 0.42 & 0.25 & 0.26 & 0.40 \\
 &  & MSE-o &  & 3.30 & 2.86 & 2.80 & 3.86 & 3.08 & 2.84 \\
 &  & MAPE-n &  & 0.64 & 0.57 & 0.65 & 0.63 & 0.67 & 0.62 \\
 &  & MAPE-o &  & 0.21 & 0.19 & 0.19 & 0.23 & 0.21 & 0.19 \\
 \hline
\multirow{4}{*}{(I)}  &  \multirow{4}{*}{0.1} & MSE-n &  & 0.26 & 0.29 & 0.42 & 0.28 & 0.28 & 0.40 \\
 &  & MSE-o &  & 2.99 & 2.76 & 2.69 & 3.01 & 2.58 & 2.73 \\
 &  & MAPE-n &  & 0.64 & 0.58 & 0.65 & 0.63 & 0.67 & 0.61 \\
 &  & MAPE-o &  & 0.20 & 0.19 & 0.19 & 0.20 & 0.19 & 0.19 \\
 \hline
\multirow{4}{*}{(II)}  &  \multirow{4}{*}{0.05} & MSE-n &  & 0.22 & 0.27 & 0.43 & 0.23 & 0.23 & 0.40 \\
 &  & MSE-o &  & 3.46 & 2.90 & 2.80 & 4.31 & 3.06 & 2.84 \\
 &  & MAPE-n &  & 0.58 & 0.52 & 0.61 & 0.57 & 0.60 & 0.58 \\
 &  & MAPE-o &  & 0.22 & 0.20 & 0.19 & 0.24 & 0.21 & 0.19 \\
 \hline
\multirow{4}{*}{(II)}  &  \multirow{4}{*}{0.1} & MSE-n &  & 0.24 & 0.28 & 0.43 & 0.27 & 0.24 & 0.40 \\
 &  & MSE-o &  & 3.05 & 2.79 & 2.78 & 3.13 & 2.60 & 2.81 \\
 &  & MAPE-n &  & 0.59 & 0.54 & 0.62 & 0.59 & 0.62 & 0.58 \\
 &  & MAPE-o &  & 0.20 & 0.19 & 0.19 & 0.20 & 0.19 & 0.19 \\
 \hline
\multirow{4}{*}{(III)}  &  \multirow{4}{*}{0.05} & MSE-n &  & 0.19 & 0.26 & 0.43 & 0.21 & 0.18 & 0.40 \\
 &  & MSE-o &  & 3.79 & 3.03 & 2.90 & 5.02 & 3.17 & 2.94 \\
 &  & MAPE-n &  & 0.50 & 0.47 & 0.57 & 0.50 & 0.48 & 0.55 \\
 &  & MAPE-o &  & 0.23 & 0.20 & 0.19 & 0.26 & 0.21 & 0.20 \\
 \hline
\multirow{4}{*}{(III)}  &  \multirow{4}{*}{0.1} & MSE-n &  & 0.22 & 0.28 & 0.44 & 0.26 & 0.20 & 0.41 \\
 &  & MSE-o &  & 3.09 & 2.78 & 2.80 & 3.25 & 2.54 & 2.82 \\
 &  & MAPE-n &  & 0.53 & 0.50 & 0.58 & 0.53 & 0.51 & 0.55 \\
 &  & MAPE-o &  & 0.20 & 0.19 & 0.19 & 0.21 & 0.19 & 0.19 \\
 \hline
\multirow{4}{*}{(IV)}  &  \multirow{4}{*}{0.05} & MSE-n &  & 0.21 & 0.27 & 0.40 & 0.21 & 0.20 & 0.40 \\
 &  & MSE-o &  & 2.38 & 1.97 & 2.01 & 2.71 & 2.52 & 2.07 \\
 &  & MAPE-n &  & 22.06 & 14.75 & 12.49 & 20.71 & 24.00 & 0.63 \\
 &  & MAPE-o &  & 0.25 & 0.22 & 0.22 & 0.27 & 0.25 & 0.22 \\
 \hline
\multirow{4}{*}{(IV)}  &  \multirow{4}{*}{0.1} & MSE-n &  & 0.23 & 0.28 & 0.42 & 0.24 & 0.23 & 0.40 \\
 &  & MSE-o &  & 2.12 & 1.95 & 2.02 & 2.14 & 2.03 & 2.07 \\
 &  & MAPE-n &  & 2.79 & 2.05 & 1.74 & 2.59 & 2.66 & 0.63 \\
 &  & MAPE-o &  & 0.23 & 0.22 & 0.22 & 0.23 & 0.21 & 0.22 \\
 \hline
\end{tabular}
\end{center}
\end{table}

%  Table
\begin{table}[htbp!]
\caption{Coverage probabilities (CP) and average lengths (AL) of $95\%$ credible intervals in non-outlying (n) and outlying (o) areas under four scenarios with $m=200$ and $\omega\in \{0.05, 0.1\}$.  
\label{tab:sim-CI}
}
\begin{center}
\begin{tabular}{ccccccccccccccc}
\hline
Scenario  & $\omega$ &  &  & IG & EH & GH & PG & & IG & EH & GH & PG\\
\hline
\multirow{4}{*}{(I)}   & \multirow{2}{*}{0.05} & n &  & 96.0 & 96.2 & 95.6 & 96.6 &  & 1.93 & 2.01 & 2.32 & 1.99 \\
 &  & o &  & 88.1 & 91.7 & 94.3 & 80.8 &  & 5.57 & 5.81 & 6.27 & 4.83 \\
 & \multirow{2}{*}{0.1} & n &  & 96.3 & 96.4 & 95.7 & 96.6 &  & 2.01 & 2.05 & 2.33 & 2.10 \\
 &  & o &  & 90.7 & 92.4 & 94.8 & 88.7 &  & 5.71 & 5.83 & 6.25 & 5.20 \\
 \hline
\multirow{4}{*}{(II)}   & \multirow{2}{*}{0.05} & n &  & 96.2 & 96.3 & 95.5 & 96.9 &  & 1.90 & 2.02 & 2.36 & 1.98 \\
 &  & o &  & 87.0 & 91.7 & 94.6 & 77.0 &  & 5.49 & 5.75 & 6.23 & 4.65 \\
 & \multirow{2}{*}{0.1} & n &  & 96.4 & 96.4 & 95.5 & 96.8 &  & 2.00 & 2.07 & 2.37 & 2.12 \\
 &  & o &  & 90.2 & 92.3 & 94.8 & 87.3 &  & 5.71 & 5.83 & 6.28 & 5.12 \\
 \hline
\multirow{4}{*}{(III)}   & \multirow{2}{*}{0.05} & n &  & 96.7 & 96.4 & 95.4 & 97.3 &  & 1.88 & 2.04 & 2.40 & 1.97 \\
 &  & o &  & 84.8 & 90.9 & 94.1 & 69.9 &  & 5.42 & 5.73 & 6.23 & 4.47 \\
 & \multirow{2}{*}{0.1} & n &  & 96.9 & 96.5 & 95.3 & 97.1 &  & 1.98 & 2.09 & 2.40 & 2.12 \\
 &  & o &  & 89.8 & 92.2 & 94.8 & 86.0 &  & 5.69 & 5.82 & 6.27 & 5.03 \\
 \hline
\multirow{4}{*}{(IV)}   & \multirow{2}{*}{0.05} & n &  & 93.9 & 95.6 & 95.4 & 95.2 &  & 1.89 & 2.01 & 2.29 & 1.91 \\
 &  & o &  & 84.5 & 91.4 & 94.3 & 77.5 &  & 4.35 & 4.83 & 5.33 & 3.80 \\
 & \multirow{2}{*}{0.1} & n &  & 94.7 & 95.8 & 95.5 & 95.7 &  & 1.99 & 2.05 & 2.33 & 2.04 \\
 &  & o &  & 88.0 & 91.6 & 94.6 & 85.8 &  & 4.51 & 4.85 & 5.32 & 4.13 \\
\hline
\end{tabular}
\end{center}
\end{table}

%----------------------------------------------------------------------------%
%         Data analysis
%----------------------------------------------------------------------------%
\section{Data analysis}
\label{sec:data}
We apply the proposed method to the analysis of crime data by the generalized linear model with Poisson likelihood and random effects. This model has been adopted for various datasets in applied statistics; examples include the modeling of areal count data in disease mapping \citep{lawson2013bayesian}. In such application, Poisson rate $\lambda_i$ (defined below) is not just an adjustment of areal effects but the parameter of interest as the intrinsic relative risk of region $i$ \citep[e.g.][]{Li2010}. 
Here we incorporate such idea of covariate adjustment into crime risk modeling.

The dataset consists of the numbers of police-recorded crime in Tokyo metropolitan area, provided by University of Tsukuba and publicly available online (``GIS database of number of police-recorded crime at O-aza, chome in Tokyo, 2009-2017'', available at \url{https://commons.sk.tsukuba.ac.jp/data_en}). 
In this study, we focus on the number of violent crimes in $m=2855$ local towns in Tokyo metropolitan area in 2015.
For auxiliary information about each town, we adopted area (km$^2$), population densities in noon and night, density of foreign people, percentage of single-person household and average duration of residence, which all help adjustment of the crime risk.
Let $y_i$ be the observed count of violent crimes, $a_i$ be area and $x_i$ be the vector of standardized auxiliary information in the $i$-th local town.
We are interested in the crime risk adjusted by the auxiliary information and, to this end, we employ the following Poisson regression model:
\begin{equation}\label{Po-crime}
y_i|\la_i\sim {\rm Po}(\la_i\eta_i), \ \ \ \ \eta_i=\exp(\log a_i+x_i^t\delta),
\end{equation}
independently for $i=1,\ldots,m$, where $\delta$ is a vector of unknown regression coefficients.
Under the model (\ref{Po-crime}), the random effect for local town $i$, $\la_i$, can be interpreted as adjustment risk factor that is not be explained by the auxiliary information.
In most local towns, the offset term explains the variations of crime rates, hence the adjustment risk factor is expected to be small. Yet, the adjustment risk might be extremely high in some local towns, and we want to detect such districts. 
This is precisely where the global-local shrinkage priors fit, for which we employed the proposed the IG and EH priors for $\la_i$. 
We adopted $N(0, 100)$ as a prior distribution of each component of $\delta$; we found the following result was robust to the choice of prior variance. 
For posterior inference, we simply use a Gibbs sampling in which the posterior samples of $\la_1,\ldots,\la_m$ and $\delta$ are iteratively drawn from their full conditional distributions. 
Conditional on $\delta$, we can still use the posterior computation algorithm for $\la_i$ provided in Section~\ref{subsec:posterior_computation}.
On the other hand, given $\lambda_i$'s, the full conditional distribution of $\delta$ is not a familiar form. The detailed algorithm customized for sampling of $\delta$ is based on the independent Metropolis-Hasting method and given in Supplementary Materials. 
For comparison, we also applied the common gamma distribution for $\la_i$ as considered in Section~\ref{sec:sim}, which is again denoted by PG in what follows.
Regarding other methods used in Section~\ref{sec:sim}, the GH prior cannot be directly applied in this case since the specification method for hyperparameters recommended in  \cite{datta2016bayesian} is reasonable only when there is no adjustment terms.
Similarly, the KW method is not applicable in this situation. Therefore, we will focus on the comparison of the proposed priors with the standard Gamma prior. 
In each Gibbs sampler, we generated 20,000 posterior samples after discarding 3,000 posterior samples as burn-in.

\begin{figure}[!htbp]
	\centering
	\includegraphics[width=15cm]{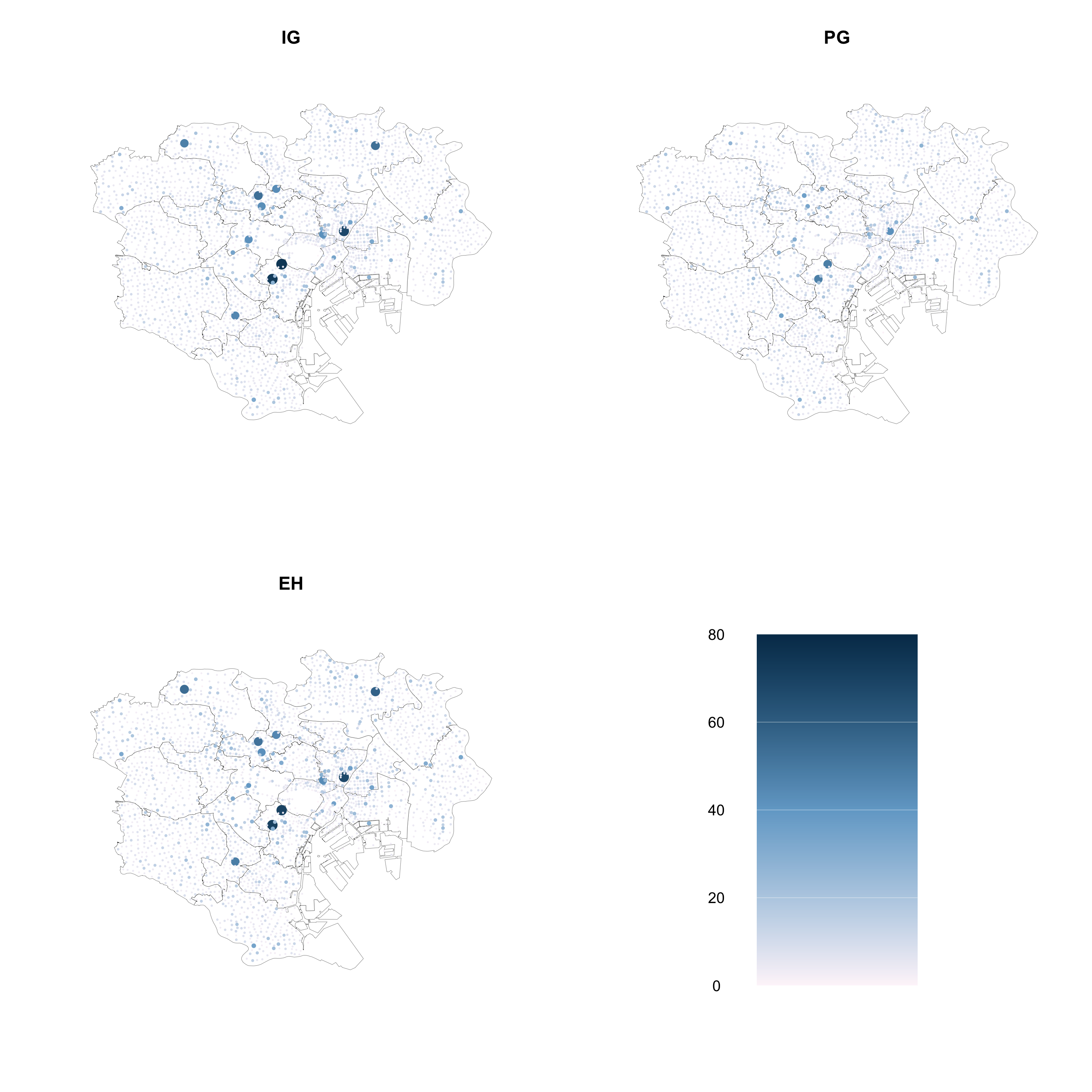}
	\caption{
	Posterior means of risk factors $\la_i$ based on IG, EH and PG methods. 
	 \label{fig:crime}
	 }
\end{figure}

\begin{figure}[!htbp]
	\centering
	\includegraphics[width=7cm]{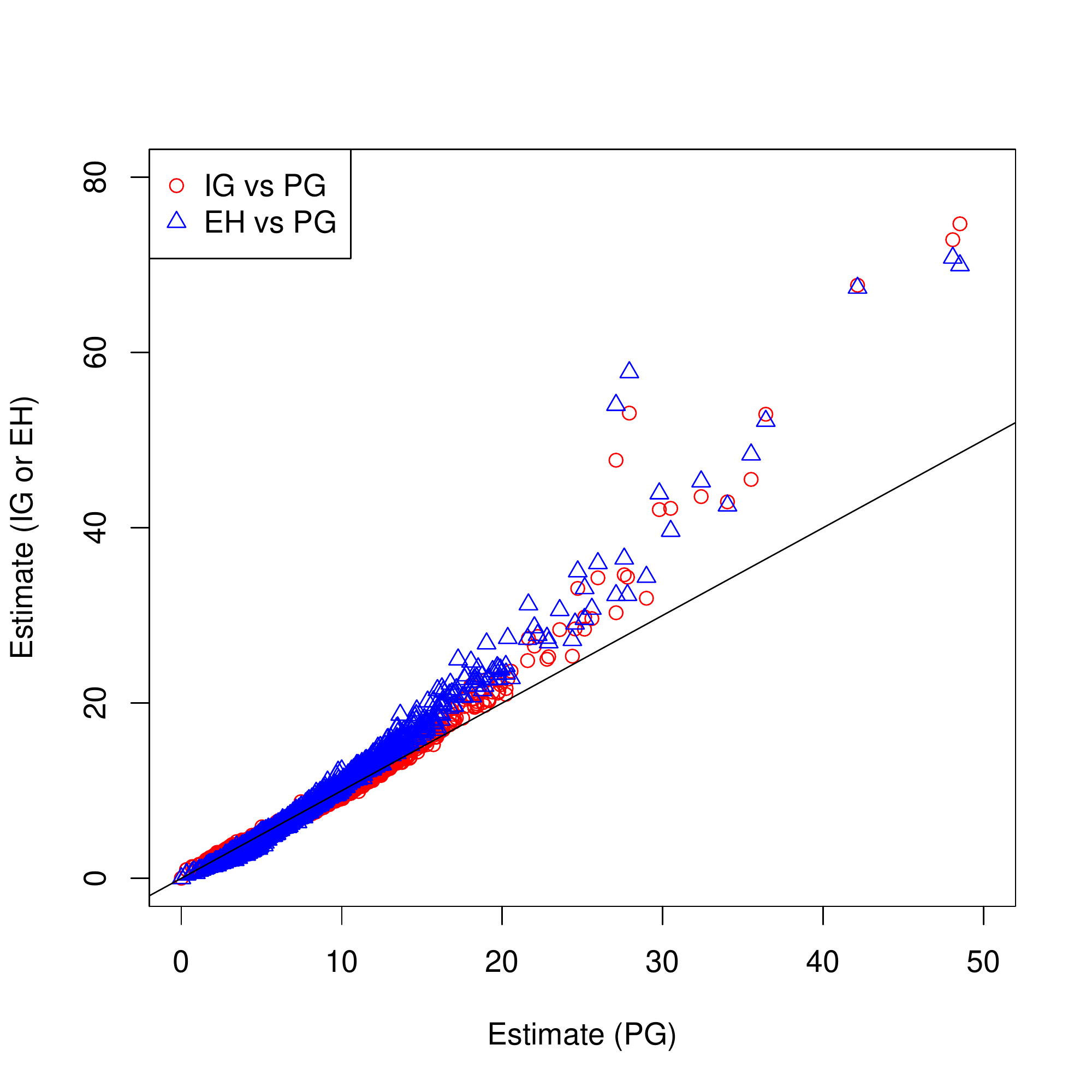}
	\caption{
	Scatter plot of posterior means of risk factors $\la_i$ based on IG, EH and PG methods. 
	 \label{fig:crime-Est}
	 }
\end{figure}

We first computed posterior means of risk factor $\lambda_i$ based on the three methods. The spatial pattern of the estimates is shown in Figure \ref{fig:crime}.
It is observed that the proposed two priors, IG and EH, produce almost the same estimates. 
We can confirm that the proposed EH method provides similar estimates of $\la_i$ in most areas and successfully detected several local towns whose risk factors are extremely high.
In contrast, such extreme towns are less emphasized, or not detected at all, by the PG method because the PG method seriously underestimates the true risk factors. 
More direct comparisons of estimates based on the purposed methods and the PG method are presented in Figure \ref{fig:crime-Est}, which indicates the underestimation property of the PG method more clearly.

\begin{figure}[!htbp]
	\centering
	\includegraphics[width=14cm]{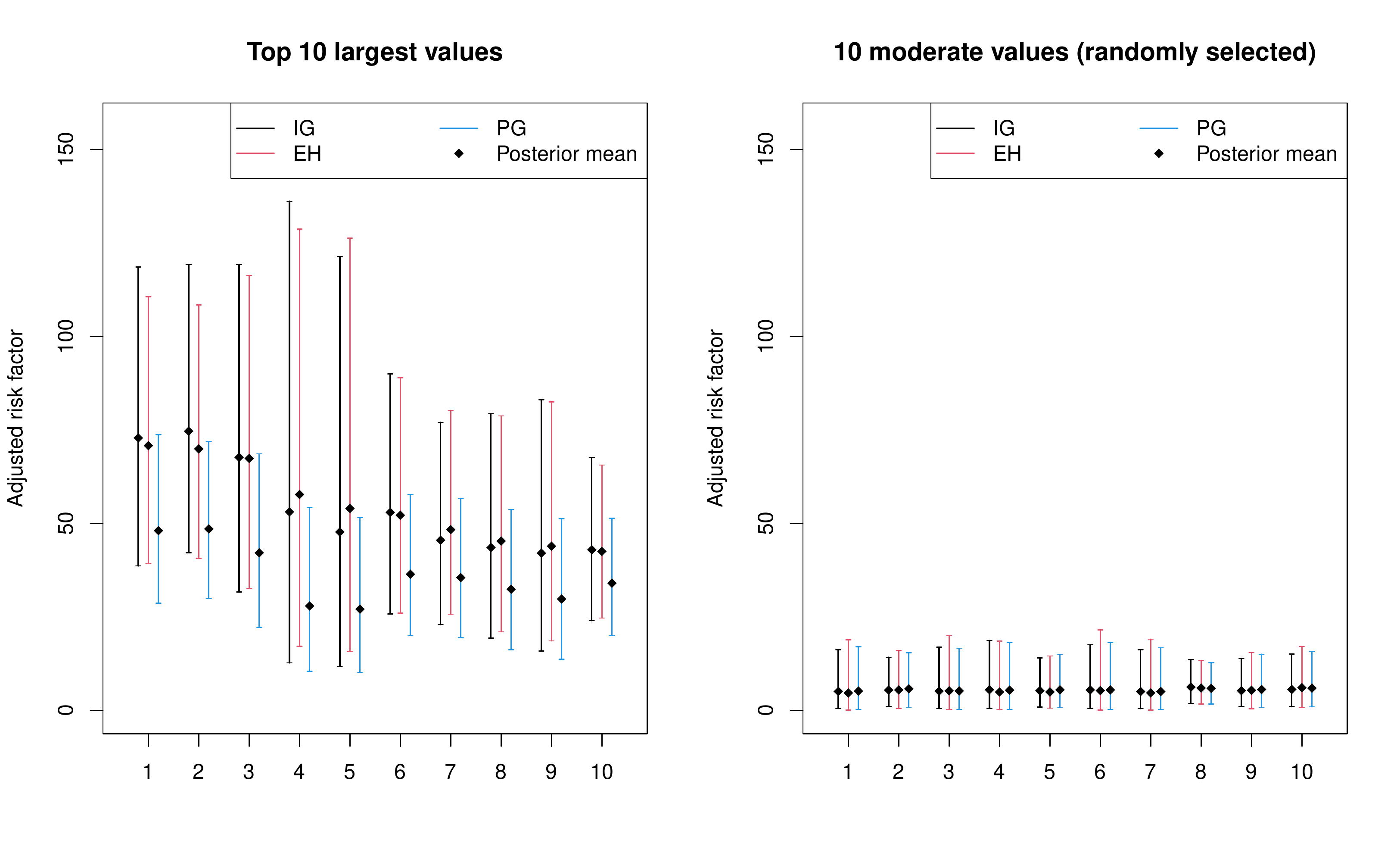}
	\caption{
	$95\%$ credible intervals for areas with highest 10 posterior means (left) and for randomly selected 10 areas with moderate posterior means (right) of adjusted risk factors.  
	 \label{fig:crime-CI}
	 }
\end{figure}

We then detected ten local towns with the largest posterior means of $\la_i$.
For these towns, we computed $95\%$ credible intervals of $\la_i$ based on the three methods, as shown in the left panel of Figure~\ref{fig:crime-CI}. 
This panel clearly shows the over-shrinkage problem of the PG method in both point estimation (posterior means) and interval estimation (posterior credible intervals); the posterior credible intervals computed by the PG method tends to be narrow and further emphasizes the underestimated results.
We also randomly selected another ten local towns with moderate estimates of $\la_i$ and gave $95\%$ credible intervals in the right panel of the same figure. 
The difference of the three methods is almost negligible for these towns. 
These observations exemplify that the proposed methods can avoid the over-shrinkage problem for large signals while their performance in the other towns are almost the same as the standard PG method.

% figure

% figure

% figure

%----------------------------------------------------------------------------%
%         Discussion
%----------------------------------------------------------------------------%
\section{Discussion}
\label{sec:dis}
It should be emphasized again that the global-local shrinkage priors for sequence of counts developed in this article are based on the new concept of tail-robustness, that is clearly different from other definitions and non-trivial for many prior densities. We provided sufficient conditions for this desirable tail-robustness property and, specifically, proposed two tractable global-local shrinkage priors. As illustrated by the simulated and real data examples, the models with these priors could actually show the tail-robustness as predicted by theory, and are expected to be applied in the studies of high-dimensional counts.

The settings of our study are critically dependent on the Poisson likelihoods, whose mean and variance are equal. Conditionally, both prior mean and variance, $\la _i$, are controlled by the common local parameter $u_i$ unde the gamma prior ${\rm Ga}(\alpha,\beta/u_i)$, affecting both the baseline of shrinkage and the amount of shrinkage. This property is not seen in the Gaussian case, where the local parameter appears in the prior variance only and controls the amount of local shrinkage, which makes the role of local parameters clear and interpretable. 
In this sense, the local parameter in (\ref{model}) might be less interpretable, while it is also this setting that enables us to carry out posterior computation easily and has been studied intensively in the literature \citep[e.g.][]{datta2016bayesian}. 
It would be an interesting future research to pursue an alternative setting for hierarchical modeling of sequence of counts under which the role of the local parameters is properly restricted and interpretable.

From the viewpoint of methodological research, this paper is primarily focused on the point and interval estimation of the Poisson rate. The high-dimensional counts can be cast as other statistical problems such as multiple testing.
The detailed investigation for such directions would extend the scope of this paper, but we leave it to a valuable future study.

The newly-introduced EH prior is motivated as the probability distributions that satisfy the conditions given in Theorem~\ref{thm:asymptotic_bias}, hence hold tail-robustness. However, the class of priors that meet those conditions is not limited to that of the EH priors. In theory, the priors with tail-robustness can be extended to 
\begin{equation*}
\pi ( u_i ) \propto \frac{ {u_i}^{\ga _1 - 1} }{  (1 + \ga _2 u_i )^{\ga _1} } \frac{1}{ \{ 1 + \ga _3 \log (1 + \ga _4 + u_i ) \} ^{1 + \ga _5} },
\end{equation*}
which is also proper and tail-robust.   
The hyperparameters $(\ga _1,\ga _2,\ga _3,\ga _4, \ga_5)$ increases the flexibility of the model and could improve the EH prior equipped with a single parameter $\ga$. 
However, the posterior inference under this prior is challenging due to the intractable normalizing constant that involves those hyperparameters. 
The full-Bayes inference for the hyperparameters is not as straightforward as that of the EH prior. The inference with fixed hyperparaemters is feasible by utilizing the same parameter augmentation in Section~\ref{subsec:posterior_computation}, but always raises the problem of hyperparameter tuning. 
We leave the development of this extension to the future work, which could be useful in more structured models for count data.

\section*{acknowledgement}
The authors are supported by the Japan Society for the Promotion of Science (JSPS KAKENHI) grant numbers: 20J10427, 17K17659 and 18K12757.

%----------------------------------------------------------------------------%
%         Sppendix
%----------------------------------------------------------------------------%
\appendix 
\section{Posterior computation algorithm}
We here provide details of the posterior computation algorithm under the proposed two priors, IG and EH priors, under the hierarchical model (\ref{model}).
The contents consists of three parts, algorithms for sampling from the common parameters in (\ref{model}), parameters related to the IG prior and parameters related to the EH prior.

\subsection*{Sampling of the common parameters $(\la _{1:m},\al,\be)$.}

\begin{itemize}
\item
The full conditional of $\la_i$ is ${\rm Ga}(y_i+\alpha,\eta_i+\beta/u_i)$ and $\la _1,\dots , \la _m$ are mutually independent. 

\item
The full conditional of $\be $ is ${\rm{Ga}} (m \al + a_{\be } , \sum_{i = 1}^{m} \la _i / u_i + b_{\be } )$. 

\item
The sampling of dispersion parameter $\al$ can be done in multiple ways. We take the strategy of \cite{zhou2013negative} by working on the conditional, negative binomial likelihood of $\alpha$ by marginalizing $\lambda _i$ out. The conditional posterior density of $\alpha$ is proportional to 
\begin{equation*}
\psi_{\al } (\alpha ) \prod_{i=1}^m \frac{\Gamma (y_i+\alpha )}{\Gamma (\alpha)} \left( \frac{\beta}{\beta + \eta_i u_i} \right) ^{\alpha} = 
\psi_{\al } (\alpha ) \prod_{i=1}^m \sum _{\nu _i = 1}^{y_i} |s(y_i,\nu _i)|\alpha ^{\nu _i} \left( 1+\frac{ \eta_i u_i}{\beta} \right) ^{-\alpha},
\end{equation*}
where $\psi_{\al } (\alpha )$ is the prior density of $\al $ and $s(y_i,\nu _i)$ is the Stirling's number of the second kind, and the summation collapses to one if $y_i=0$. The integer-valued variable $\nu _i$ is considered a latent parameter that augments the model and allows Gibbs sampler. Thus, we need to sample from the full conditionals of $\alpha$ and $\nu _{1:m}$.

\begin{itemize}
\item 
The conditional of $\al$ is ${\rm{Ga}} ( \sum _{i=1}^m \nu _i + a_{\alpha} , \sum _{i=1}^m \log (1+\eta_iu_i/\beta ) + b_{\alpha} )$. 

\item 
If $y_i=0$, then $\nu _i = 0$ with probability one. Otherwise, the conditional posterior probability function of $\nu _i$ is proportional to $|s(y_i,\nu _i)|\alpha ^{\nu _i}$, from which we can sample based on the distributional equation $\nu _i = \sum _{j=1}^{y_i} d_j$, where $d_j \ (j=1,\ldots,y_i)$ are independent random variables distributed as ${\rm{Ber}} \left( \alpha/(j-1+\alpha) \right)$.
      
\end{itemize}
\end{itemize}

\subsection*{Sampling of parameters related to IG prior}
\begin{itemize}
\item
The full conditional of $u_i$ is $\mathrm{IG}(\ga + \al , \ga + \la _i\be )$ and $u_1,\dots , u_m$ are mutually independent. 

\item
The full conditional of $\ga $ is proportional to 
$$
f_{\ga } ( \ga ) = {\ga ^{m \ga } \over \{ \Ga ( \ga ) \} ^m} \Big( \prod_{i = 1}^{m} {1 \over u_i} \Big) ^{\ga } \exp\left(- \ga \sum_{i = 1}^{m}\frac{1}{u_i}\right) I(\ep_1\leq \gamma \leq \ep_2) \text{.} \non
$$ 
We sample the candidate of $\ga$, denoted ${\ga }^{*}$, from the distribution of $\min\{\ep_2, \max \{ \ep_1 , Z \}\} $, $Z \sim {\rm{N}} ( \widetilde{\ga } , \si ^2 )$, with tuning parameter $\si > 0$, where $\widetilde{\ga } $ is the current value of $\ga $, and accept it with probability $\min \{ 1, f_{\ga } ( {\ga }^{*} ) / f_{\ga } ( \tilde{\ga } ) \} $, where we assume that the correction factor based on the asymmetric proposal density can be ignored. 
\end{itemize}

\subsection*{Sampling parameters related to the EH prior}

The latent parameters, $(v_i,w_i)$, are marginalized out except for the sampling of $u_i$ (Partially collapsed Gibbs sampler, \citealt{van2008partially}). 

\begin{itemize}
 
\item
The full conditional of $u_i$ is ${\rm{GIG}} (1 - \al , 2v_i, 2 \be \la _i )$, where ${\rm GIG}(a,b,p)$ is the generalized inverse Gaussian distribution with density $\pi(x; a,b,p)\propto x^{p-1} \exp \{ -( a x + b x)/2 \}$ for $x>0$. 

\item
The full conditional of $(v_i|w_i)$ is ${\rm{Ga}} (1 + w_i, 1 + u_i)$. The conditional posterior of $w_i$ with $v_i$ marginalized out is ${\rm{Ga}} (1 + \ga , 1 + \log (1 + u_i ))$. 

\item 
Under gamma prior $\gamma\sim {\rm Ga}(a_{\gamma},b_{\gamma})$, the full conditional of $\ga$ (with $(v_i,w_i)$ marginalized out) is ${\rm{Ga}} (a_{\ga }+m , b_{\ga}+\sum_{i = 1}^{m} \log \{ 1 + \log (1 + u_i ) \})$. 

\end{itemize}

%  Reference
\vspace{5mm}
\bibliographystyle{chicago}
\bibliography{CountShrinkage}

\begin{thebibliography}{}

\bibitem[\protect\citeauthoryear{Armagan, Clyde, and Dunson}{Armagan
  et~al.}{2011}]{armagan2011generalized}
Armagan, A., M.~Clyde, and D.~B. Dunson (2011).
\newblock Generalized beta mixtures of gaussians.
\newblock In {\em Advances in neural information processing systems}, pp.\
  523--531.

\bibitem[\protect\citeauthoryear{Berry, Helman, and West}{Berry
  et~al.}{2019}]{BerryWest2018TSM}
Berry, L.~R., P.~Helman, and M.~West (2019).
\newblock Probabilistic forecasting of heterogeneous consumer transaction-sales
  time series.
\newblock arXiv:1808.04698.

\bibitem[\protect\citeauthoryear{Bhadra, Datta, Polson, and Willard}{Bhadra
  et~al.}{2016}]{bhadra2016default}
Bhadra, A., J.~Datta, N.~G. Polson, and B.~Willard (2016).
\newblock Default bayesian analysis with global-local shrinkage priors.
\newblock {\em Biometrika\/}~{\em 103\/}(4), 955--969.

\bibitem[\protect\citeauthoryear{Bhadra, Datta, Polson, and Willard}{Bhadra
  et~al.}{2019}]{bhadra2019lasso}
Bhadra, A., J.~Datta, N.~G. Polson, and B.~T. Willard (2019).
\newblock Lasso meets horseshoe: A survey.
\newblock {\em Statistical Science, to appear\/}.

\bibitem[\protect\citeauthoryear{Brown, Greenshtein, and Ritov}{Brown
  et~al.}{2013}]{brown2013poisson}
Brown, L.~D., E.~Greenshtein, and Y.~Ritov (2013).
\newblock The poisson compound decision problem revisited.
\newblock {\em Journal of the American Statistical Association\/}~{\em
  108\/}(502), 741--749.

\bibitem[\protect\citeauthoryear{Carvalho, Polson, and Scott}{Carvalho
  et~al.}{2010}]{carvalho2010horseshoe}
Carvalho, C.~M., N.~G. Polson, and J.~G. Scott (2010).
\newblock The horseshoe estimator for sparse signals.
\newblock {\em Biometrika\/}~{\em 97\/}(2), 465--480.

\bibitem[\protect\citeauthoryear{Datta and Dunson}{Datta and
  Dunson}{2016}]{datta2016bayesian}
Datta, J. and D.~B. Dunson (2016).
\newblock Bayesian inference on quasi-sparse count data.
\newblock {\em Biometrika\/}~{\em 103\/}(4), 971--983.

\bibitem[\protect\citeauthoryear{Kiefer and Wolfowitz}{Kiefer and
  Wolfowitz}{1956}]{KW1956}
Kiefer, J. and J.~Wolfowitz (1956).
\newblock Consistency of the maximum likelihood estimator in the presence of
  infinitely many incidental parameters.
\newblock {\em Annals of Mathematical Statistics\/}~{\em 27}, 887--906.

\bibitem[\protect\citeauthoryear{Koenker and Mizera}{Koenker and
  Mizera}{2014}]{koenker2014convex}
Koenker, R. and I.~Mizera (2014).
\newblock Convex optimization, shape constraints, compound decisions, and
  empirical bayes rules.
\newblock {\em Journal of the American Statistical Association\/}~{\em
  109\/}(506), 674--685.

\bibitem[\protect\citeauthoryear{Lawson}{Lawson}{2013}]{lawson2013bayesian}
Lawson, A.~B. (2013).
\newblock {\em Bayesian disease mapping: hierarchical modeling in spatial
  epidemiology}.
\newblock Chapman and Hall/CRC.

\bibitem[\protect\citeauthoryear{Li, Graubard, and Gail}{Li
  et~al.}{2010}]{Li2010}
Li, H., B.~I. Graubard, and M.~H. Gail (2010).
\newblock Covariate adjustment and ranking methods to identify regions with
  high and low mortality rates.
\newblock {\em Biometrics\/}~{\em 66}, 613--620.

\bibitem[\protect\citeauthoryear{Maruyama and Strawderman}{Maruyama and
  Strawderman}{2019}]{maruyama2019admissible}
Maruyama, Y. and W.~E. Strawderman (2019).
\newblock Admissible bayes equivariant estimation of location vectors for
  spherically symmetric distributions with unknown scale.
\newblock {\em Annals of Statistics, to appear.\/}.

\bibitem[\protect\citeauthoryear{Seneta}{Seneta}{1976}]{senetaregularly}
Seneta, E. (1976).
\newblock {\em Regularly varying functions}, Volume 508.
\newblock Springer-Verlag Berlin Heidelberg.

\bibitem[\protect\citeauthoryear{Tang, Ghosh, Ha, and Sedransk}{Tang
  et~al.}{2018}]{tang2018modeling}
Tang, X., M.~Ghosh, N.~S. Ha, and J.~Sedransk (2018).
\newblock Modeling random effects using global--local shrinkage priors in small
  area estimation.
\newblock {\em Journal of the American Statistical Association\/}~{\em
  113\/}(524), 1476--1489.

\bibitem[\protect\citeauthoryear{van Dyk and Park}{van Dyk and
  Park}{2008}]{van2008partially}
van Dyk, D.~A. and T.~Park (2008).
\newblock Partially collapsed gibbs samplers: Theory and methods.
\newblock {\em Journal of the American Statistical Association\/}~{\em 103},
  790--796.

\bibitem[\protect\citeauthoryear{Wakefield}{Wakefield}{2006}]{wakefield2006disease}
Wakefield, J. (2006).
\newblock Disease mapping and spatial regression with count data.
\newblock {\em Biostatistics\/}~{\em 8\/}(2), 158--183.

\bibitem[\protect\citeauthoryear{Yano, Kaneko, and Komaki}{Yano
  et~al.}{2019}]{yano2019}
Yano, K., R.~Kaneko, and F.~Komaki (2019).
\newblock Exact minimax predictive density for sparse count data.
\newblock arXiv:1812.06037.

\bibitem[\protect\citeauthoryear{Zhou and Carin}{Zhou and
  Carin}{2013}]{zhou2013negative}
Zhou, M. and L.~Carin (2013).
\newblock Negative binomial process count and mixture modeling.
\newblock {\em IEEE Transactions on Pattern Analysis and Machine
  Intelligence\/}~{\em 37\/}(2), 307--320.

\bibitem[\protect\citeauthoryear{Zhu, Ibrahim, and Love}{Zhu
  et~al.}{2019}]{Zhu2019}
Zhu, A., J.~G. Ibrahim, and M.~I. Love (2019).
\newblock Heavy-tailed prior distributions for sequence count data: removing
  the noise and preserving large differences.
\newblock {\em Bioinformatics\/}~{\em 35\/}(12), 2084--2092.

\end{thebibliography}

\newpage
%----------------------------------------------------------------------------%
%     Supplementary Material
%----------------------------------------------------------------------------%
\begin{center}
{\LARGE \bf Supplemental Materials for ``On Global-local Shrinkage Priors for Count Data''}
\end{center}

%  new environment
\setcounter{equation}{0}
\renewcommand{\theequation}{S\arabic{equation}}
\setcounter{section}{0}
\renewcommand{\thesection}{S\arabic{section}}
\setcounter{table}{0}
\renewcommand{\thetable}{S\arabic{table}}
\setcounter{page}{1}

\vspace{1cm}
In this Supplementary Material, we provide technical details regarding the proofs of Theorem~1, Proposition~1, motivation of the EH prior and other computational issues in the main text.

%----------------------------------------------------------------------------%
%        Lemmas
%----------------------------------------------------------------------------%
\section{Lemmas}

In this section, we provide two lemmas which will be used in the proof of Theorem 1 in the next section. 
The following lemma is useful for proving Proposition 1 as well.

\begin{lem}
	\label{Lem:convergence}
	Let $0 < M_1 < M_0 < \infty $. 
	Let $h_0 ( \cdot )$ and $h_1 ( \cdot )$ be nonnegative integrable functions defined on $( M_0 , \infty )$ and $(0, M_1 )$, respectively, and let $0 < \varphi ( \cdot ) < 1$ be a strictly increasing function defined on $(0, \infty )$. 
	Suppose that $\int_{M_0}^{\infty } h_0 (u) du > 0$. 
	Then 
	\begin{align}
	\lim_{y \to \infty } \int_{0}^{M_1} \{ \varphi (u) \} ^y h_1 (u) du / \int_{M_0}^{\infty } \{ \varphi (u) \} ^y h_0 (u) du = 0 \text{.} \non
	\end{align}
\end{lem}

\noindent
{\bf Proof.} \ \ 
We have 
\begin{align}
&\limsup_{y \to \infty } \int_{0}^{M_1} \{ \varphi (u) \} ^y h_1 (u) du / \int_{M_0}^{\infty } \{ \varphi (u) \} ^y h_0 (u) du \non \\
&\le \limsup_{y \to \infty } \Big\{ \frac{\varphi ( M_1 ) }{ \varphi ( M_0 )} \Big\} ^y \int_{0}^{M_1} h_1 (u) du / \int_{M_0}^{\infty } h_0 (u) du \non \\
&= 0 \non
\end{align}
by assumption. 
\hfill$\Box$

\begin{lem}
\label{lem:S1S2} 
The assumptions of Theorem 1 imply the following: 
\begin{align}
%&\int_{0}^{\infty } \frac{\pi (u) }{ ( \be + u)^{\al }} d u + \int_{0}^{\infty } \frac{|u {\pi }' (u)| }{ ( \be + u)^{\al }} d u < \infty \text{,} \label{A1_1} \\
&\int_{0}^{\infty } \frac{|u {\pi }' (u)| }{ ( \be + u)^{\al }} d u < \infty \text{,} \label{A1_1} \\
&\lim_{u \to \infty } \frac{u \pi (u) }{ ( \be + u)^{\al }} = \lim_{u \to 0} \frac{u \pi (u) }{ ( \be + u)^{\al }} = 0 \text{.} \label{A1_2} 
\end{align}
\end{lem}

\noindent
{\bf Proof.} \ \ %We first show that the assumptions imply the following: 
We first note that if $\pi ( \cdot )$ is to be proper, we must have $\xi \in [- \infty , 0]$ since otherwise $\pi ( \cdot )$ would be eventually increasing so that 
\begin{align}
\int_{N}^{\infty } \pi (u) du \ge \int_{N}^{\infty } \pi (N) du = \infty \non 
\end{align}
for some $N > 0$. 
%The first term of (\ref{A1_1}) is finite since $\pi ( \cdot )$ is proper. 
By (A1) of the main text, we have 
\begin{align}
\int_{0}^{1} \frac{|u {\pi }' (u)| }{ ( \be + u)^{\al }} du \le \int_{0}^{1} {|u {\pi }' (u)| \over \be ^{\al }} du < \infty \text{.} \non 
\end{align}
If $\xi > - \infty $, then $|u \pi{}' (u)| = O( \pi (u))$ as $u \to \infty $ and hence 
\begin{align}
\int_{1}^{\infty } \frac{|u {\pi }' (u)| }{ ( \be + u)^{\al }} du \le \int_{1}^{\infty } {|u {\pi }' (u)| \over \be ^{\al }} du < \infty \text{.} \non
\end{align}
On the other hand, if $\xi = - \infty $, then there exists $N > 0$ such that $\pi{}' (u) < 0$ for all $u \ge N$ and therefore 
\begin{align}
\infty &> \limsup_{M \to \infty } \int_{N}^{M} \pi (u) du = \limsup_{M \to \infty } \Big[ M \pi (M) - N \pi (N) + \int_{N}^{M} \{ - u {\pi }' (u) \} du \Big] \non \\
&\ge \limsup_{M \to \infty } \Big[ - N \pi (N) + \int_{N}^{M} \{ - u {\pi }' (u) \} du \Big] = - N \pi (N) + \int_{N}^{\infty } |u {\pi }' (u)| du \non 
\end{align}
by integration by parts. 
Thus, (\ref{A1_1}) follows. 
To prove (\ref{A1_2}), note that for any $0 < \de < 1 < M < \infty $, we have 
\begin{align}
\Big[ u \pi (u) \Big] _{\de }^{M} &= \int_{\de }^{M} \pi (u) du + \int_{\de }^{M} u {\pi }' (u) du \non 
\end{align}
by integration by parts. 
Then, since the right-hand side of the above equation converges as $\de \to 0$ and as $M \to \infty $, there exist $c_0 , \mathring{c} \in [0, \infty)$ such that $\lim_{u \to 0} u \pi (u) = c_0$ and $\lim_{u \to \infty } u \pi (u) = \mathring{c}$. 
If $c_0 > 0$ or $\mathring{c} > 0$, then $u^{- 1} = O( \pi (u))$ as $u \to 0$ or as $u \to \infty $ in contradiction to the assumption that $\pi ( \cdot )$ is proper. 
Thus, $c_0 = \mathring{c} = 0$ and (\ref{A1_2}) follows. 
\hfill$\Box$

%----------------------------------------------------------------------------%
%         Proof of Theorem 1
%----------------------------------------------------------------------------%
\section{Proof of Theorem~1}

In this section, we prove the following theorem. 

\begin{thm}
Assume that $\pi ( \cdot )$ is strictly positive and continuously differentiable. 
Suppose that $\pi ( \cdot )$ satisfies the following two conditions: 
\begin{align}
&\int_{0}^{1} |u \pi{}' (u)| du < \infty \text{,} \tag{A1} \label{A_1} \\
&\xi \equiv \lim_{u \to \infty } \frac{u {\pi }' (u) }{ \pi (u)} \quad \text{exists} \ \text{in} \  [- \infty , \infty ] \text{.} \tag{A2} \label{A_2}
\end{align}
Then the asymptotic bias of $\lat _i$ is $1 + \xi $, that is, 
\begin{align}
\lim_{y_i \to \infty } ( \lat _i - y_i ) = 1 + \xi \text{.} \non
\end{align}
\end{thm}

\noindent
{\bf Proof.} \ \ We prove the result by using (\ref{A1_1}), (\ref{A1_2}), and (A2). 
Since the posterior density of $u_i$ given $y_i$ is proportional to $W( u_i ) \pi ( u_i )$,
where $W ( u_i ) = W( u_i ; y_i ) = {u_i}^{y_i} / (1 + u_i / \be )^{y_i + \al }$, the difference between $y_i$ and $\lat _i$ is 
\begin{align}
y_i  - \lat _i 
&= \int_{0}^{\infty } \frac{y_i - \al u_i / \be }{ 1 + u_i / \be } W(u_i)\pi(u_i) d u_i \bigg/
\int_{0}^{\infty } W(u_i)\pi(u_i) d u_i  \text{,} \label{tuip0}
\end{align}
which is finite by the propriety of the posterior. 
By making the change of variables $t = ( u_i / \be ) / (1 + u_i / \be )$, we have 
\begin{align}
y_i  - \lat _i 
=  \int_{0}^{1} \{ y_i - ( y_i + \al ) t \} g(t) dt \bigg/ \int_{0}^{1} g(t) dt  \text{,} \non
\end{align}
where $g(t) = g(t; y_i ) = t^{y_i} (1 - t)^{\al - 2} \pi ( \be t / (1 - t))$. 
Note that, by integration by parts and (\ref{A1_2}), 
\begin{align}
( \al + 1) \int_{0}^{1} t g(t) dt &= \int_{0}^{1} ( \al + 1) (1 - t)^{\al } t^{y_i + 1} (1 - t)^{- 2} \pi \Big( \be \frac{t}{1 - t} \Big) dt \non \\
&= \Big[ - (1 - t)^{\al + 1} t^{y_i + 1} (1 - t)^{- 2} \pi \Big( \be \frac{t}{1 - t} \Big) \Big] _{t = 0}^{t = 1} \non \\
&\quad + \int_{0}^{1} \Big[ (1 - t)^{\al + 1} t^{y_i + 1} (1 - t)^{- 2} \pi \Big( \be \frac{t}{1 - t} \Big) \non \\
&\quad \quad \times \Big\{ {y_i + 1 \over t} + {2 \over 1 - t} + {\partial \over \partial t} \log \pi \Big( \be \frac{t}{1 - t} \Big) \Big\} \Big] dt \non \\
&= ( y_i + 1) \int_{0}^{1} (1 - t) g(t) dt + 2 \int_{0}^{1} t g(t) dt \non \\
&\quad + \be \int_{0}^{1} \frac{t}{1 - t} \Big\{ {\pi }' \Big( \be \frac{t}{1 - t} \Big) / \pi \Big( \be \frac{t}{1 - t} \Big) \Big\} g(t) dt \text{,} \non
\end{align}
or 
\begin{align}
\int_{0}^{1} \{ y_i - ( y_i + \al ) t \} g(t) dt = - \int_{0}^{1} g(t) dt - \be \int_{0}^{1} \frac{t}{1 - t} \Big\{ {\pi }' \Big( \be \frac{t}{1 - t} \Big) / \pi \Big( \be \frac{t}{1 - t} \Big) \Big\} g(t) dt \text{.} \non
\end{align}
Then, by making the change of variables $u_i = \be t / (1 - t)$, we obtain 
\begin{align}
y_i  - \lat _i &= - 1 - \be \left(\int_{0}^{1} g(t) dt\right)^{-1}\int_{0}^{1} \frac{t}{1 - t} \Big\{ {\pi }' \Big( \be \frac{t}{1 - t} \Big) / \pi \Big( \be \frac{t}{1 - t} \Big) \Big\} g(t) dt  \non \\
&= - 1 - \int_{0}^{\infty } H(u_i) u_i {\pi }' ( u_i )  d u_i \bigg/ \int_{0}^{\infty } H(u_i) \pi ( u_i ) d u_i, \non
\end{align}
where the integrands are absolutely integrable by (\ref{A1_1}) and 
$$
H(u_i)=H(u_i;\beta)=\left( \frac{u_i / \be}{1 + u_i / \be} \right) ^{y_i}\frac{1}{(1 + u_i / \be )^{\al }}.
$$

Now, suppose first that $\xi > - \infty $. 
Then, for any $M > 0$, 
\begin{align}
&| y_i - \lat _i + 1 + \xi | \\
&\le  \int_{0}^{\infty } \Big| \xi - \frac{u_i {\pi }' ( u_i ) }{ {\pi } ( u_i )} \Big| H(u_i) \pi ( u_i ) d u_i 
\bigg/
\int_{0}^{\infty } H(u_i) \pi ( u_i ) d u_i \\ 
&= \frac{ \int_{0}^{\infty } H(u_i) h_{1} ( u_i ) d u_i }{ \int_{0}^{\infty } H(u_i) h_{0} ( u_i ) d u_i } \non \\
&\le \frac{ \int_{0}^{M} H(u_i) h_{1} ( u_i ) d u_i }{ \int_{M + 1}^{\infty } H(u_i) h_{0} ( u_i ) d u_i } + \frac{ \int_{M}^{\infty } H(u_i) h_{1} ( u_i ) d u_i }{ \int_{0}^{\infty } H(u_i) h_{0} ( u_i ) d u_i } \text{,} \non 
\end{align}
where $h_{k} ( u_i ) = | \xi - u_i {\pi }' ( u_i ) / {\pi } ( u_i )| ^k {\pi } ( u_i )$ for $k = 0, 1$. 
The first term in the fourth line converges to zero as $y_i \to \infty $ by Lemma \ref{Lem:convergence}. 
On the other hand, 
\begin{align}
&\limsup_{M \to \infty } \sup_{y_i \in \{ 0, 1, 2, \dotsc \} } \frac{ \int_{M}^{\infty } H(u_i) h_{1} ( u_i ) d u_i }{ \int_{0}^{\infty } H(u_i) h_{0} ( u_i ) d u_i } \non \\
&= \limsup_{M \to \infty } \sup_{y_i \in \{ 0, 1, 2, \dotsc \} } \frac{ \int_{M}^{\infty } \big| \xi - u_i \pi' ( u_i ) / \pi ( u_i ) \big| H(u_i) h_{0} ( u_i ) d u_i }{ \int_{M}^{\infty } H(u_i) h_{0} ( u_i ) d u_i } \non \\
&\le \limsup_{M \to \infty } \sup_{u_i \in (M, \infty )} \Big| \xi - \frac{u_i {\pi }' ( u_i ) }{\pi ( u_i )} \Big| = \lim_{u_i \to \infty } \Big| \xi - \frac{u_i {\pi }' ( u_i ) }{\pi ( u_i )} \Big| = 0 \text{.} \non 
\end{align}
Thus, 
\begin{align}
\limsup _{y_i \to \infty } | y_i  - \lat _i + 1 + \xi | &\le \limsup _{M \to \infty } \limsup _{y_i \to \infty } \frac{ \int_{0}^{M} H(u_i) h_{1} ( u_i ) d u_i }{ \int_{M + 1}^{\infty } H(u_i) h_{0} ( u_i ) d u_i } \non \\
&\quad + \limsup _{M \to \infty } \limsup _{y_i \to \infty } \frac{ \int_{M}^{\infty } H(u_i) h_{1} ( u_i ) d u_i }{ \int_{0}^{\infty } H(u_i) h_{0} ( u_i ) d u_i } \non \\
&\le 0 + 0 = 0 \text{.} \non 
\end{align}

Next, suppose that $\xi = - \infty $. 
Then for any $M > 0$, there exists $N > 0$ such that $- u \pi{}' (u) / \pi (u) > M$ for all $u \ge N$. 
Therefore, 
\begin{align}
y_i  - \lat _i + 1 
&= - \frac{ \int_{0}^{N} H(u_i) u_i {\pi }' ( u_i ) d u_i }{ \int_{0}^{\infty } H(u_i) \pi ( u_i ) d u_i }
-\frac{ \int_{N}^{\infty } H(u_i) u_i {\pi }' ( u_i )  d u_i }{ \int_{0}^{\infty } H(u_i) \pi ( u_i ) d u_i } \non \\
&\ge - \frac{ \int_{0}^{N} H(u_i) u_i {\pi }' ( u_i )  d u_i }{ \int_{0}^{\infty } H(u_i) \pi ( u_i ) d u_i }
+ M \frac{ \int_{N}^{\infty } H(u_i) \pi ( u_i )d u_i }{ \int_{0}^{\infty } H(u_i) \pi ( u_i ) d u_i } \to M \non 
\end{align}
as $y_i \to \infty $ by Lemma \ref{Lem:convergence}. 
Thus, since $M$ is arbitrary, we conclude that 
\begin{align}
\lim_{y_i \to \infty } ( \lat _i - y_i ) = - \infty = 1 + \xi \text{.} \non 
\end{align}
This completes the proof.  
\hfill$\Box$

%----------------------------------------------------------------------------%
%         Other tail robustness 
%----------------------------------------------------------------------------% 
\section{Related tail-robustness properties} 

We here discuss two related tail-robustness properties of the posterior mean $\lat_i$.
One variant is based on the ratio of the estimator and observation and given by 
\begin{align}
\lim_{y_i \to \infty } {| \lat _i - y_i | \over y_i} = 0 \text{,} 
\label{eq:tail_robustness_weak} 
\end{align}
which we name {\it weak tail-robustness}. 
It is obvious that the strong tail-robustness implies the weak one. The left-hand-side in (\ref{eq:tail_robustness_weak}) is the mean absolute percentage error (MAPE) loss function, which is frequently used in practice to evaluate the inferential/predictive performance of the models for count data. In this sense, the weakly tail-robust estimator $\lat _i$ is asymptotically optimal in MAPE (Section 3.3.2, \cite{BerryWest2018TSM}). 
Note that the Bayes estimator $( \al + y_i ) / (1 + \be / u_i )$ with fixed $u_i$ does not satisfy the property (\ref{eq:tail_robustness_weak}).

We provide conditions for weak tail-robustness in the following proposition. 
\begin{prp}
Suppose that $\pi ( \cdot )$ is strictly positive. 
Then, under the model (2.1),
we have 
\begin{align}
\lim_{y_i \to \infty } {| \lat _i - y_i | \over y_i} = 0 \text{.} \non 
\end{align}
i.e., the Bayes estimator is weakly tail-robust. 
\end{prp}

\noindent
{\bf Proof.} \ \ From (\ref{tuip0}), we have 
\begin{align}
{\lat _i - y_i \over y_i} &= - \frac{ \be \int_{0}^{\infty } (\be + u_i)^{-1} W(u_i)\pi(u_i) d u_i }{ \int_{0}^{\infty } W(u_i)\pi(u_i) d u_i } + \frac{\al }{y_i} \frac{ \int_{0}^{\infty } u_i(\beta+u_i)^{-1} W(u_i)\pi(u_i) d u_i }{ \int_{0}^{\infty } W(u_i)\pi(u_i) d u_i } \label{pMAPEp1}
\end{align}
for $y_i \in \{ 1, 2, \dotsc \} $, where 
$$
W(u_i)=W(u_i; y_i )=\frac{{u_i}^{y_i}}{ (1 + u_i / \be )^{y_i + \al }}.
$$
The second term on the right-hand side of (\ref{pMAPEp1}) converges to zero as $y_i \to \infty $ since $u_i/(\beta+u_i) \le 1$ for all $u_i \in (0, \infty )$. 
On the other hand, by Lemma \ref{Lem:convergence}, 
\begin{align}
&\frac{  \int_{0}^{\infty } \be(\beta+u_i)^{-1} W(u_i)\pi(u_i) d u_i }{ \int_{0}^{\infty } W(u_i)\pi(u_i) d u_i } \\
& \ \ = \frac{ \int_{M}^{\infty } \be(\beta+u_i)^{-1} W(u_i)\pi(u_i) d u_i }{ \int_{M}^{\infty } W(u_i)\pi(u_i) d u_i } \non \\
&\quad \times \frac{ \int_{0}^{\infty } \be(\beta+u_i)^{-1} W(u_i)\pi(u_i) / \int_{M}^{\infty } \be(\beta+u_i)^{-1} W(u_i)\pi(u_i) d u_i }{ \int_{0}^{\infty } W(u_i)\pi(u_i) d u_i / \int_{M}^{\infty } W(u_i)\pi(u_i) d u_i } \non \\
& \ \ \sim \frac{ \int_{M}^{\infty } \be(\beta+u_i)^{-1} W(u_i)\pi(u_i) d u_i }{ \int_{M}^{\infty } W(u_i)\pi(u_i) d u_i } \non
\end{align}
as $y_i \to \infty $ for every $M > 0$. 
Furthermore, uniformly in $y_i$, 
\begin{align}
0 \le \frac{ \int_{M}^{\infty } \be(\beta+u_i)^{-1} W(u_i)\pi(u_i) d u_i }{ \int_{M}^{\infty } W(u_i)\pi(u_i) d u_i } \le {\be \over \be + M  } \to 0 \quad \text{as} \quad M \to \infty \text{.} \non
\end{align}
Thus, we have proved the desired result. 
\hfill$\Box$
\bigskip

The key implication of this proposition is that, for fixed hyperparameters, the weak tail-robustness can be achieved for almost all priors for $u_i$. 
It suggests that the tail-robustness property in the main document and the weak tail-robustness property look similar but is substantially different properties.

Another concept of tail-robustness is 
\begin{align}
\lim_{y_i \to \infty } {\lat_i \over \al + y_i} = 1 \text{.} 
\label{eq:tail_robustness_DattaDunson}
\end{align}
The denominator is a part of the Bayes estimator $( \al + y_i ) / (1 + \be / u_i )$.
This definition requires that the coefficient $u_i/(\be + u_i)$, viewed as a shrinkage factor, degenerates at $1$ as $y_i\to \infty$ (Proposition~1, \citealt{datta2016bayesian}). 
It is trivial that the Bayes estimator with fixed $u_i$ does not satisfy the property, but the weak tail-robustness leads to the tail-robustness of this type.
Hence, by Proposition~1, the use of any strictly positive prior of $u_i$ also leads to this tail-robustness.

%----------------------------------------------------------------------------%
%         Derivation of EH
%----------------------------------------------------------------------------%
\section{Connection to the tail-robustness of three-parameter beta priors}

The EH prior emerges in the course of examination of tail-robustness under the scaled-beta or three-parameter beta (TPB) distributions \citep{armagan2011generalized}, known as a flexible class of priors for scale parameters.
The density is given by 
\begin{align}
\pi ( u_i ; a_0,b_0,\phi_0,\ga_0 ) \propto \frac{{u_i}^{a_0 - 1}}{(1 + \phi _0 u_i / \be )^{\ga _0} (1 + u_i / \be )^{a_0 + b_0 - \ga _0}} \text{,} \label{eq:prior_TPB}
\end{align}
where $a_0$, $b_0$, $\phi _0$, and $\ga _0$ are all positive constants. For count data and Poisson likelihood, \cite{datta2016bayesian} considered this prior with $a_0 = b_0 = 1 / 2$ and $\phi _0 = \be $. 
Although the TPB prior (\ref{eq:prior_TPB}) is flexible, it does not satisfy assumption (A3) in Corollary~1 and is not strongly tail-robust for any choice of hyperparameters. 
Under the prior (\ref{eq:prior_TPB}), by Theorem 1, the asymptotic bias is $\lim_{y_i \to \infty } ( \lat _i - y_i ) = - b_0 < 0$, negatively biased and dependent on its hyperparameter $b_0$. 
Similar to the inverse-gamma prior, the approximate tail-robustness for the TPB prior is justified by the limiting case of $b_0 \to 0$; with $\ga _0 = \ga $, $a_0 = 1 + \ga $, and $\be = \phi _0 = 1$ in (\ref{eq:prior_TPB}), we obtain 
\begin{equation}
\pi( u_i ; \ga ) \propto \frac{u_i^\gamma}{(1+u_i)^{1+\gamma}}, \ \ \ \gamma >  - 1 \text{.} \label{eq:improper_prior} 
\end{equation}
In return for the tail-robustness in the limit, however, it is inevitable for the prior in (\ref{eq:improper_prior}) to be improper. The EH prior can be viewed as the limit $\ga \to \infty$ modified by the multiplied log-term for propriety.

%----------------------------------------------------------------------------%
%         Computation of the marginal prior
%----------------------------------------------------------------------------%
\section{Evaluation of the marginal of $\la _i$ with EH prior}

We evaluate the limit of the marginal density of $\la_i$ implied by the EH prior. 
\begin{align}
{p( \la _i ; \al , \be , \ga ) \over \pi_{\rm{EH}} ( \la _i ; \ga )} &= {\be ^{\al } \over \Ga ( \al )} \int_{0}^{\infty } {e^{- \be / x} \over x^{\al }} {1 + \la _i \over 1 + \la _i x} \Big\{ {1 + \log (1 + \la _i ) \over 1 + \log (1 + \la _i x)} \Big\} ^{1 + \ga } dx \non 
\end{align}
To compute the limit at $\la _i = \infty$, note that 
\begin{align}
\lim _{\la _i\to \infty} {1 + \la _i \over 1 + \la _i x} \Big\{ {1 + \log (1 + \la _i ) \over 1 + \log (1 + \la _i x)} \Big\} ^{1 + \ga } = \frac{1}{x} \non 
\end{align}
for each $x>0$. The result in the main text is verified by the dominated convergence theorem. To see this, evaluate the integrand for $\la _i \ge 1$ as 
\begin{align}
{1 + \la _i \over 1 + \la _i x} \Big\{ {1 + \log (1 + \la _i ) \over 1 + \log (1 + \la _i x)} \Big\} ^{1 + \ga } &\le {2 \over x} \exp \Big( (1 + \ga ) \Big[ \log \{ 1 + \log (1 + \la _i s) \} \Big] _{s = x}^{s = 1} \Big) \non \\
&= {2 \over x} \exp \Big\{ (1 + \ga ) \int_{x}^{1} \frac{1}{s} {\la _is \over 1 + \la _i s} {1 \over 1 + \log (1 + \la _i s)} ds \Big\} \non \\
&\le \begin{cases} \displaystyle {2 \over x} & \text{if $x \ge 1$} \\ \displaystyle {2 \over x} \exp \Big\{ (1 + \ga ) \int_{x}^{1} {1 \over s} ds \Big\} & \text{if $x < 1$} \end{cases} \non \\
&\le {2 \over x} + {2 \over x} \exp \Big\{ (1 + \ga ) \int_{x}^{1} {1 \over s} ds \Big\} = {2 \over x} \Big( 1 + {1 \over x^{1 + \ga }} \Big) \non 
\end{align}
in which we find the bounding function that is integrable as 
\begin{align}
\int_{0}^{\infty } {e^{- \be / x} \over x^{\al }} {2 \over x} \Big( 1 + {1 \over x^{1 + \ga }} \Big) dx < \infty \non 
\end{align}
for large $\la _i > 1$.

%----------------------------------------------------------------------------%
%         Additional simulation results
%----------------------------------------------------------------------------%
\section{Additional simulation results}
We here provide additional simulation results under a larger sample size ($m=400$), where the other settings are the same as ones in the main document. 
The results are shown in Tables \ref{tab:sim} and \ref{tab:sim-CI}. 
We can see that the results are not very different from Tables 2 and 3 in the main document.

We also assessed the computation time of the proposed methods, IG and EH, and two existing Bayesian methods, GH and PG. 
Using scenario (I) given in the main document with $\omega=0.1$, we evaluated the computation time under $m\in \{200, 400\}$. 
For each $m$, 3000 posterior samples were generated after discarding the first 500 burn-in samples. 
The computation time is reported in Table \ref{tab:CT}, where the experiment was performed on a PC with 3.2 GHz 8-Core Intel Xeon W 8 Core Processor with approximately 32GB RAM. 
From Table \ref{tab:CT}, we can see that the computation time of the proposed methods, IG and EH, are quite comparable with that of PG, and are considerably smaller than that of GH.
Moreover, as the number of (local) parameters in the four models linearly increase with $m$, their computation time would also linearly increase with $m$, which is partly supported by the results in Table \ref{tab:CT}.

%  Table
\begin{table}[htbp!]
\caption{Averaged values of mean squared errors (MSE) and mean absolute percentage error (MAPE) in non-outlying (-n) and outlying (-o) areas under four scenarios with $m=400$ and $\omega\in \{0.05, 0.1\}$.  
\label{tab:sim}
}
\begin{center}
\begin{tabular}{ccccccccccccccc}
\hline
Scenario  & $\omega$ &  &  & IG & EH & GH & PG & KW & ML \\
\hline
\multirow{4}{*}{(I)} & \multirow{4}{*}{0.05} & MSE-n &  & 0.24 & 0.27 & 0.42 & 0.25 & 0.25 & 0.40 \\
 &  & MSE-o &  & 3.29 & 2.93 & 2.80 & 3.88 & 2.88 & 2.85 \\
 &  & MAPE-n &  & 0.64 & 0.57 & 0.65 & 0.62 & 0.67 & 0.61 \\
 &  & MAPE-o &  & 0.21 & 0.20 & 0.19 & 0.23 & 0.20 & 0.19 \\
\hline \multirow{4}{*}{(I)} & \multirow{4}{*}{0.1} & MSE-n &  & 0.26 & 0.28 & 0.43 & 0.28 & 0.27 & 0.40 \\
 &  & MSE-o &  & 3.05 & 2.85 & 2.80 & 3.02 & 2.51 & 2.84 \\
 &  & MAPE-n &  & 0.65 & 0.59 & 0.66 & 0.63 & 0.68 & 0.61 \\
 &  & MAPE-o &  & 0.20 & 0.19 & 0.19 & 0.20 & 0.19 & 0.19 \\
\hline \multirow{4}{*}{(II)} & \multirow{4}{*}{0.05} & MSE-n &  & 0.21 & 0.26 & 0.43 & 0.23 & 0.22 & 0.40 \\
 &  & MSE-o &  & 3.40 & 2.92 & 2.77 & 4.35 & 2.85 & 2.80 \\
 &  & MAPE-n &  & 0.59 & 0.53 & 0.61 & 0.57 & 0.60 & 0.58 \\
 &  & MAPE-o &  & 0.21 & 0.20 & 0.19 & 0.24 & 0.20 & 0.19 \\
\hline  \multirow{4}{*}{(II)} & \multirow{4}{*}{0.1} & MSE-n &  & 0.23 & 0.28 & 0.43 & 0.27 & 0.24 & 0.40 \\
 &  & MSE-o &  & 3.09 & 2.83 & 2.80 & 3.17 & 2.46 & 2.83 \\
 &  & MAPE-n &  & 0.59 & 0.54 & 0.62 & 0.58 & 0.61 & 0.58 \\
 &  & MAPE-o &  & 0.20 & 0.19 & 0.19 & 0.20 & 0.18 & 0.19 \\
\hline \multirow{4}{*}{(III)} & \multirow{4}{*}{0.05} & MSE-n &  & 0.19 & 0.25 & 0.43 & 0.21 & 0.17 & 0.40 \\
 &  & MSE-o &  & 3.56 & 2.94 & 2.76 & 4.87 & 2.84 & 2.80 \\
 &  & MAPE-n &  & 0.50 & 0.48 & 0.58 & 0.51 & 0.49 & 0.55 \\
 &  & MAPE-o &  & 0.22 & 0.20 & 0.19 & 0.26 & 0.20 & 0.19 \\
\hline \multirow{4}{*}{(III)} & \multirow{4}{*}{0.1} & MSE-n &  & 0.21 & 0.27 & 0.44 & 0.26 & 0.19 & 0.40 \\
 &  & MSE-o &  & 3.14 & 2.83 & 2.80 & 3.34 & 2.41 & 2.82 \\
 &  & MAPE-n &  & 0.52 & 0.49 & 0.58 & 0.53 & 0.49 & 0.55 \\
 &  & MAPE-o &  & 0.21 & 0.19 & 0.19 & 0.21 & 0.18 & 0.19 \\
\hline \multirow{4}{*}{(IV)} & \multirow{4}{*}{0.05} & MSE-n &  & 0.21 & 0.26 & 0.40 & 0.21 & 0.20 & 0.40 \\
 &  & MSE-o &  & 2.38 & 1.97 & 2.00 & 2.67 & 2.37 & 2.07 \\
 &  & MAPE-n &  & 2.38 & 1.65 & 1.36 & 2.19 & 2.14 & 0.63 \\
 &  & MAPE-o &  & 0.25 & 0.22 & 0.22 & 0.26 & 0.24 & 0.22 \\
\hline \multirow{4}{*}{(IV)} & \multirow{4}{*}{0.1} & MSE-n &  & 0.23 & 0.27 & 0.42 & 0.24 & 0.23 & 0.40 \\
 &  & MSE-o &  & 2.09 & 1.91 & 1.99 & 2.10 & 1.90 & 2.04 \\
 &  & MAPE-n &  & 2.42 & 1.77 & 1.45 & 2.19 & 2.30 & 0.63 \\
 &  & MAPE-o &  & 0.23 & 0.22 & 0.22 & 0.23 & 0.20 & 0.22 \\
\hline
\end{tabular}
\end{center}
\end{table}

%  Table
\begin{table}[htbp!]
\caption{Coverage probabilities (CP) and average lengths (AL) of $95\%$ credible intervals in non-outlying (n) and outlying (o) areas under four scenarios with $m=400$ and $\omega\in \{0.05, 0.1\}$.  
\label{tab:sim-CI}
}
\begin{center}
\begin{tabular}{ccccccccccccccc}
\hline
Scenario  & $\omega$ &  &  & IG & EH & GH & PG & & IG & EH & GH & PG\\
\hline  \multirow{4}{*}{(I)}   & \multirow{2}{*}{0.05} & n &  & 95.7 & 96.3 & 95.7 & 96.7 &  & 1.91 & 2.00 & 2.33 & 1.99 \\
 &  & o &  & 88.4 & 91.0 & 94.5 & 80.7 &  & 5.61 & 5.72 & 6.25 & 4.81 \\
 & \multirow{2}{*}{0.1} & n &  & 96.1 & 96.5 & 95.7 & 96.6 &  & 1.99 & 2.05 & 2.34 & 2.10 \\
 &  & o &  & 90.6 & 92.0 & 94.7 & 88.3 &  & 5.76 & 5.78 & 6.26 & 5.19 \\
\hline  \multirow{4}{*}{(II)}   & \multirow{2}{*}{0.05} & n &  & 95.8 & 96.4 & 95.5 & 96.9 &  & 1.88 & 2.02 & 2.36 & 1.98 \\
 &  & o &  & 87.6 & 90.9 & 94.6 & 76.4 &  & 5.56 & 5.67 & 6.25 & 4.63 \\
 & \multirow{2}{*}{0.1} & n &  & 96.2 & 96.6 & 95.5 & 96.8 &  & 1.97 & 2.07 & 2.37 & 2.12 \\
 &  & o &  & 90.1 & 91.6 & 94.7 & 86.8 &  & 5.76 & 5.78 & 6.29 & 5.11 \\
\hline  \multirow{4}{*}{(III)}   & \multirow{2}{*}{0.05} & n &  & 96.2 & 96.5 & 95.3 & 97.3 &  & 1.85 & 2.03 & 2.40 & 1.97 \\
 &  & o &  & 85.9 & 90.4 & 94.8 & 70.9 &  & 5.49 & 5.64 & 6.24 & 4.45 \\
 & \multirow{2}{*}{0.1} & n &  & 96.5 & 96.6 & 95.4 & 97.1 &  & 1.95 & 2.09 & 2.40 & 2.12 \\
 &  & o &  & 89.8 & 91.7 & 94.8 & 85.5 &  & 5.73 & 5.75 & 6.28 & 5.00 \\
\hline  \multirow{4}{*}{(IV)}   & \multirow{2}{*}{0.05} & n &  & 93.7 & 95.6 & 95.4 & 95.2 &  & 1.88 & 2.00 & 2.29 & 1.92 \\
 &  & o &  & 84.2 & 90.8 & 94.5 & 78.3 &  & 4.31 & 4.73 & 5.30 & 3.79 \\
 & \multirow{2}{*}{0.1} & n &  & 94.5 & 95.8 & 95.5 & 95.8 &  & 1.98 & 2.05 & 2.33 & 2.03 \\
 &  & o &  & 88.2 & 91.7 & 94.6 & 86.5 &  & 4.51 & 4.79 & 5.33 & 4.13 \\

\hline
\end{tabular}
\end{center}
\end{table}

%  Table
\begin{table}[htbp!]
\caption{Computation time (seconds) of the four Bayesian methods with $m=200$ and $m=400$. 
In all the methods, 3000 posterior samples were generated after discarding the first 500 samples. 
\label{tab:CT}
}
\begin{center}
\begin{tabular}{ccccccccccccccc}
\hline
& $m$  &  & IG & EH & GH & PG \\
\hline
\multirow{2}{*}{Computation Time} & $200$ & & 2.00 & 5.49 & 19.24 & 1.75 \\
& $400$ & & 3.92 & 11.06 & 38.18 & 3.65 \\
\hline
\end{tabular}
\end{center}
\end{table}

\section{Metropolis-Hastings method for Poisson regression}

The estimation of Poisson regression model in Section~5 requires the sampling of regression coefficients $\delta$, in addition to $\lambda _i$ and other parameters. The new step of sampling $\delta$ is added to the existing MCMC algorithm in Appendix, as described here. 

Consider the conditionally independent counts $y_1,\dots, y_m$ that follow
\begin{equation*}
y_i \sim Po(\lambda _i \eta _i), \ \ \ \ \ \ \ \eta _i = \exp \{ x_i'\delta \} ,
\end{equation*}
where $\lambda_i$ is a random, individual effect, $x_i$ is the $p$-vector of covariates and $\delta$ is the coefficient vector. If the likelihood has a known offset term $a_i$ as $y_i\sim Po(a_i\lambda _i \eta _i)$, then read $\lambda_i$ in the equation above as $a_i\lambda_i$. We are interested in the posterior analysis of $(\lambda _{1:m} , \delta )$ (and the other parameters) by Gibbs sampler. For $\lambda _i$, the gamma prior is conditionally conjugate; if $\lambda_i \sim Ga(\alpha , \beta /u_i)$, then the conditional posterior of $\lambda_i$ is $Ga(\alpha + y_i , \beta/u_i + \eta _i)$. With offset $a_i$, the conditional posterior is $Ga(\alpha + y_i , \beta/u_i + a_i\eta _i)$. The sampling of the other parameters is not affected by the introduction of regression and offset terms. In this note, we explain the sampling of $\delta$ by MCMC method. 

The independent Metropolis-Hastings method can be tailored for the model with conditional posterior density that is analytically available or, at least, numerically evaluated. For the Poisson regression model, we assume the normal prior $N(\mu_0 ,\Sigma _0)$ for $\delta$. Conditional on $\lambda _{1:m}$ and current $\delta ^{old}$, we generate the candidate $\delta ^{new}$ from the proposal distribution, which is defined as the posterior distribution derived from the approximate likelihood,
\begin{equation*}
\hat{\delta} \sim N(\delta,\hat{\Sigma}),
\end{equation*} 
for some known $\hat{\delta}$ and $\hat{\Sigma}$. Then, denote $\eta _i^{new} = \exp \{ x_i'\delta ^{new} \}$ and $\eta _i^{old} = \exp \{ x_i'\delta ^{old} \}$, and accept $\delta ^{new}$ with probability 
\begin{equation} \label{acc}
\min \left\{ \ 1, \ \prod _{i=1}^m \frac{ Po(y_i| \lambda _i \eta _i^{new}) N(\hat{\delta} | \delta ^{old} , \hat{\Sigma} ) }{Po(y_i| \lambda _i \eta _i^{old}) N(\hat{\delta} | \delta ^{new}, \hat{\Sigma} )} \ \right\}
\end{equation}
and set $\delta = \delta ^{new}$. Otherwise, set $\delta = \delta ^{old}$. 

The approximate normal likelihood is obtained as the Taylor expansion of the log-likelihood around the mode. The log-likelihood of this model is  
\begin{equation*}
\begin{split}
\ell (\delta ) &= \sum _{i=1}^m \log \left( \frac{\lambda _i^{y_i}}{y_i!} \right) + y_i \log ( \eta _i ) - \lambda _i \eta _i \\
&= \mathrm{const.} + \sum _{i=1}^m y_i (x_i'\delta ) - \lambda _i e^{ x_i'\delta }  
\end{split}
\end{equation*}
The first and second derivatives are
\begin{equation*}
\frac{\partial \ell (\delta ) }{\partial \delta} = \sum _{i=1}^m y_i x_i - \lambda _i e^{ x_i'\delta } x_i \ \ \ \ \ \ \ \ \ \mathrm{and} \ \ \ \ \ \ \ \ \ \frac{\partial \ell (\delta ) }{\partial \delta \partial \delta'} = - \sum _{i=1}^m \lambda _i e^{ x_i'\delta } x_ix_i' 
\end{equation*}
Then, we obtain $\hat{\delta}$ as the solution of the first order condition, 
\begin{equation} \label{first}
\frac{\partial \ell (\delta ) }{\partial \delta} = 0_p, \ \ \ \ \ \ \ \ \mathrm{i.e.,} \ \ \ \ \ \ \ \ \ \ \sum _{i=1}^m y_i x_i = \sum _{i=1}^m \lambda _i e^{ x_i'\delta } x_i
\end{equation}
where $0_p$ is the $p$-vector of zeros. The precision is obtained by 
\begin{equation} \label{second}
\hat{\Sigma}^{-1} = \left. - \frac{\partial \ell (\delta ) }{\partial \delta \partial \delta'} \right| _{\delta = \hat{\delta}} = \sum _{i=1}^m \lambda _i e^{ x_i' \hat{\delta} } x_ix_i' 
\end{equation}
The computation of $\hat{\delta}$ needs the numerical solver of the nonlinear equation above. It should be noted that we {\it do not have to solve this equation exactly}, for the solution $\hat{\delta}$ is used to construct the approximate, proposal distribution. The sampling from the proposal is justified by the acceptance-rejection step, no matter what the proposal distribution is used. 

In summary, the sampling of $\delta$ takes the following steps. Conditional on $\delta^{old}$ and the other parameters,
\begin{enumerate}
    \item Compute $\hat{\delta}$ and $\hat{\Sigma}$ by Equations~(\ref{first}) and (\ref{second}). 
    
    \item Generate $\delta ^{new}$ from the proposal distribution $N(\mu ,\Sigma)$, where \begin{equation*}
        \Sigma = ( \hat{\Sigma}^{-1} + \Sigma_0^{-1} )^{-1}, \ \ \ \ \ \ \mu = \Sigma ( \hat{\Sigma}^{-1}\hat{\delta} + \Sigma_0^{-1}\mu _0 ).
    \end{equation*}
    
    \item Set $\delta = \delta ^{new}$ with probability given in Equation (\ref{acc}). Otherwise, set $\delta = \delta^{old}$. 
    
\end{enumerate}

\end{document}